\documentclass[a4paper,12pt]{article}
\usepackage{jheppub}
\usepackage{mathtools}
\usepackage{amssymb,amstext}
\usepackage{graphicx}
\usepackage{epsfig}
\usepackage[stable]{footmisc}
\usepackage[utf8]{inputenc}
\usepackage{subcaption}
\usepackage{lscape}

\newcommand{\cno}[1]{\,\textbf{:}#1\textbf{:}\,}

\newcommand{\N}{{\cal N}}
\newcommand{\Op}{{\cal O}}
\newcommand{\D}{{\mathfrak D}}
\newcommand{\R}{{\mathcal R}}

\newcommand{\K}{{\mathcal K}}
\newcommand{\LL}{{\mathcal L}}

\newcommand{\dD}{{\delta\mathfrak D}}

\newcommand{\W}{{\mathcal W}}
\newcommand{\spc}{~,\qquad}

\DeclareMathOperator{\SU}{\mathrm{SU}}

\DeclareMathOperator{\psu}{\mathfrak{psu}}

\DeclareMathOperator{\su}{\mathfrak{su}}

\DeclareMathOperator{\ssl}{\mathfrak{sl}}

\newcommand{\pd}{\partial}

\newcommand{\fdf}[2]{\frac{\delta #1}{\delta #2}}

\DeclareMathOperator{\tr}{Tr}

\newcommand{\bra}[1]{\left\langle~#1~\right\vert}
\newcommand{\ket}[1]{\left\vert~ #1 ~\right\rangle}

\newcommand{\zbra}[1]{\Bigl\langle\relax{\kern-.4em}\Bigl\langle~#1~\Bigr\vert}
\newcommand{\zinnp}[2]{\Bigl\langle\relax{\kern-.4em}\Bigl\langle~ #1~\Bigm\vert~ #2~\Bigr\rangle}            %Zamolodchikov's inner product
\newcommand{\zbok}[3]{\Bigl\langle\relax{\kern-.4em}\Bigl\langle ~#1~\Bigl\vert~ #2~\Bigr\vert~#3~\Bigr\rangle} %Zamolodchikov's expectation value

\begin{document}

\begin{flushright}
WIS/07/14-AUG-DPPA
\end{flushright}

%%%%%%%%%%  TITLE AND ABSTRACT 

\begin{center} {\Large \bf Chiral 2D "Strange Metals" from ${\cal N}=4$ SYM}
\end{center} 
%\vskip 1cm   

\vspace{7mm}

\centerline{Micha Berkooz\footnote{\tt email: micha.berkooz@weizmann.ac.il}, Prithvi Narayan\footnote{\tt email:  prithvi@weizmann.ac.il} and Amir Zait\footnote{\tt email: amir.zait@weizmann.ac.il}}

\vspace{5mm}

\bigskip\centerline{\it Department of Particle Physics and Astrophysics, Weizmann Institute of Science,}
\smallskip\centerline{\it  Rehovot 76100, Israel}

\bigskip\medskip
%\vfil

%%%%%%%%%

\begin{abstract}
\ Familiar field theories may contain closed subsectors made out of only fermions, which can be used to explore new and unusual phases of matter in lower dimensions. We focus on the fermionic $\su(1,1)$ sector in ${\cal N}=4$ SYM and on its ground states, which are Fermi surface states/operators. By computing their spectrum to order $(g_{YM}^2 N)^2$, we argue that fluctuations around this fermi surface, within the sector and in the limit $k_F\rightarrow\infty$, are governed by a chiral 1+1 dimensional sector of the "strange metal" coset $SU(N)_N \otimes SU(N)_N/SU(N)_{2N}$. On the gravity side, the conjectured dual configuration is an $S=0$ degeneration of a rotating black hole. On general grounds we expect that the near horizon excitations of $(S=0,\Omega=1,J\rightarrow\infty)$ degenerations of black holes will be governed by a chiral sector of a 1+1 CFT.
 \end{abstract}

%%%%%%%%%  END OF TITLE AND ABSTRACT

\section{Introduction}

The $AdS/CFT$ correspondence \cite{Maldacena:1997re,Witten:1998qj,Gubser:2002tv} is a powerful tool used to obtain results both on strongly coupled field theories and on String theory or gravity. In particular, it has led to a much better understanding of black holes to the extent that they can be enumerated qualitatively in most cases, and precisely in some. We would like to extend our understanding to additional kinds of black holes, and new phases of String theory which arise at their near horizon limits. 

The claim that we will make in this paper is that certain small fluctuations of Fermi surface states in the $\psu(1,1|2)$ and fermionic $\su(1,1)$ sectors of ${\cal N}=4$ SYM are governed by a chiral sector (say, right moving) of the 1+1 "strange metal" \cite{Gopakumar}  gauged model 
\begin{equation}\label{coset}
\frac{SU(N)_N \otimes SU(N)_N}{ SU(N)_{2N} }
\end{equation} 
The origin of the numerator will be the fermions in ${\cal N}=4$ SYM. To argue for the gauging, we will compute, at weak coupling to order $g^4$ at large $N$(where $g$ is 't~Hooft coupling), the spectrum of these fluctuations and demonstrate how the gauging comes about dynamically at low energies. Supersymmetry does not play an essential role in these arguments as the Fermi surface states are not supersymmetric to start with\footnote{Although they are close to them in a sense which will be made clear below.}.

When taken in conjunction with the conjectured duality \cite{Dori,Micha} between these Fermi surface states and a specific class of singular degenerations of black holes in $AdS_5\times S^5$ \cite{Mei2007,Chong:2005da,Chong:2006zx}  (or more precisely, within a familiar consistent truncation of the latter), we conclude that the near horizon of these black holes contains a sector governed by a higher spin integrable CFT. I.e., we can provide a workable example of a higher spin theory (albeit chiral) within theories that we are familiar with, and a flow which interpolates the latter to the former.

\medskip

If true, then this construction might shed light on several aspects of black holes physics and the $AdS/CFT$ duality:

{\bf 1. Theories with $W$-symmetries within String theory}: The tensionless limit of string theory has recently emerged as a promising new example of the AdS/CFT correspondence \cite{Klebanov:2002ja}. In this duality, the bulk side contains an infinite tower of massless higher spin fields which enhance the gauge symmetry, which can then be employed to gain insights into the working of the AdS/CFT duality. In this work we will be interested in 2 dimensional CFTs, which are possibly dual to 3 dimensional bulk higher spin theories. In the context of two dimensional CFTs these enhanced symmetries are usually termed $W$ symmetries (see \cite{Bouwknegt}). CFTs which are vector-like (i.e. their central charge is $c \sim N$ in the large-$N$ limit) with $W_N$ symmetries have been proposed as duals to Vasiliev higher spin theories on $AdS_3$ \cite{Gaberdiel:2010,Gaberdiel:2012}. In this work we encounter CFTs which are matrix like (with central charge $c \sim N^2$) with "extended" $W_N$ symmetries. Our results suggest that for the chiral sector of these CFTs the bulk dual is a string theory on Near horizon geometries of certain fast rotating black holes in an otherwise familiar and benign $AdS$ space. This gives an explicit realization of theories with $W$ symmetry in CFTs within a known bulk dual.

{\bf 2. Singular degenerations of black holes} Since we start in a familiar $AdS$ space, in this case $AdS_5$, we do not have the freedom to tune the string tension to zero. Rather, we go to the tensionless limit by having some curvature diverge somewhere within the solution, which is the case for the $S=0$ black hole that we have. 

These black holes are, however, degenerations of otherwise reasonable black holes, i.e., by slightly heating up the system we go to a black hole with $S>0$, for which the horizon is smooth and with low curvature. Reversing the argument, we start with these $S>0$ black holes and go to the extremal limit, where $S\rightarrow 0$ as well. When this happens the horizon recedes, shrinks and collapses around a ring of singularities which is now naked. We will refer to the final configurations as a singular degenerations of a black holes. Bulk computations can be carried at $S>0$ and then one can try and extrapolate them to the singular limit, relying on some intuition from the dual field theory to address potential problems of instability. 

Furthermore, such degenerations can be obtained in various dimensions and are in no way unique to $AdS_5$. Therefore they might teach us new lessons on a larger class of singularities in string theory.

{\bf 3. Applying integrability techniques to the study of black holes} Techniques borrowed from integrable systems have proven to be essential in understanding the string worldsheet in some $AdS$ spaces, and the spectrum of excited strings. Understanding black holes is outside the scope of such techniques both because of their different large-$N$ scaling, and, perhaps more critically, because fast scrambling systems such as black holes are not described by integrable systems. 

This is the case for a general black hole, but it may be better for our specific class of extremal black holes. The $S=0$ black hole is conjectured to be made out of partons in a specific subsector of ${\cal N}=4$ SYM - the fermionic $\psu(1,1)$ sector, which can be embedded in the $\psu(1,1|2)$ sector\footnote{These sectors may also be related to the problem of classifying low SUSY operators in ${\cal N}=4,2$ theories \cite{Micha,Dori}} \cite{Zwiebel,Beisert2007}. In fact, it is the ground state in this sector (for a given charge). The corresponding operator is therefore dual to an exact eigenstate of the dilatation operator \cite{Dori}, and it is an interesting open question whether one can compute its dimensions to all order in perturbation theory. 

In any case, the conjecture that some of the near-horizon fluctuations are given by the coset in (\ref{coset}) implies that the near horizon should have again a familiar large $N$ limit, even though these are excitations about a state which is far from being a long trace operator.

{\bf 4. New phases of electronic matter:} The $AdS/CFT$ duality has been used to obtain insight into strongly correlated electron systems in condensed matter systems \cite{Hartnoll,Faulkner}. For example, the duality is very useful in taking into account the dynamics of order parameters in large N theories across the entire RG flow, and extremal black hole configurations have proven to be useful in such setups. Since the ground state of most CM systems is non-degenerate, it is natural to combine the latter with an $S\rightarrow 0$ limit, as is the case for our dual black holes. Such degenerations may exist in other known $AdS/CFT$ pairs and we expect that the specific model $SU(N)_N \otimes SU(N)_N/SU(N)_{2N}$ can be generalized to other 1+1 CFTs. The model, however, is intimately tied to the 1+1 dimensional chiral nature of the construction, and hence it is not clear if one will be able to generalize it to higher dimensions or to  non-chiral case. What is likely, however, is that any $S=0$ degeneration of an extremal black hole are a natural starting point for applications to CM systems.

{\bf 5. $(S=0,\Omega=1,J\rightarrow\infty)$ Black holes and chiral sectors of CFTs}

The construction that we will present is closely tied to the fact that the black holes are fast rotating with $\Omega\rightarrow 1$. In the operator language, the field theory dual will contain a restricted number of "letters" from the field theory dictionary, and only a single derivative $\partial_{1{\dot 1}}$. When going to the picture of states on an $S^3\times R_t$, in radial quantization, these correspond to quanta moving along a "large circle" of the $S^3$. Consider a general $S^d$ and a state which has large angular momenta along a fixed two plane which intersects this $S^d$. Generally, particles with high momenta along this circle, need not be governed by a local theory on this circle since (virtual) particles can "make a short cut" from one point of the circle to another via the rest of the $S^d$.  However, in the limit of very high momenta along this circle we expect the theory to become local because (1) it started its life a local theory on the sphere, and (2) at large momenta all emissions are boosted to a very narrow cone around the circle. I.e., particles are kinematically confined to the "big circle" and do not transverse the sphere away from it.

The conclusion is that if we take any black hole, in any $AdS_d, \ d>3$ space, and spin it to the limit of $\Omega\rightarrow 1$ along a two-plane, then the theory will be governed by a chiral sector of a local 1+1 field theory, i.e., states with conformal dimensions $(0,h)$ in a local 1+1 CFT. If we further truncate to an $S=0$ configuration, then we are in the ground state of such a theory. Operators with dimensions $(0,h)$ are either Virasoro descendants of the identity operator or else constitute an extended chiral symmetry of some sort, such as $W$-symmetry.

\medskip

The discussion above might also be related to some recent discussion in the literature: 

1. Using the $\psu(1,1|2)$ sector of $\mathcal{N}=4$ SYM is advantageous 
for our purposes since it is a non-compact sector, allowing for rich, though still controlled, dynamics. Recently, a sector with similar symmetry algebra and field content was used in the context of four  dimensional ${\cal N}=2$ SCFTs \cite{Beem:2013sza} in order to derive strong restrictions on their spectrum. In that work, 
%By restricting a $d>2$ CFT to a single plane, one could naively hope to be able to use the powerful methods of two-dimensional conformal symmetry in order to obtain exact results for the original theory. As it turns out, this insight cannot be used in the general case by picking $\ssl(2) \times \bar{\ssl(2)} \in \so(d+2)$, since all chiral operators with respect to one of the $\ssl(2)$ factors are in fact trivial under $\so(d+2)$ (NOT CLEAR TO ME). As shown in \cite{Beem:2013sza}, however, the situation improves dramatically in $\mathcal N = 2$ SCFTs, 
one can find a twisted Virasoro subalgebra in a restricted 2-plane in 4D, which is non-trivial in the full theory, 
composed of both the $SU(2)_R$ symmetry and a subgroup of the Poincar\"e symmetry. 
This maps a subsector of the four dimensional theory, which has a $\psu(1,1|2)$ algebra
to a (non-unitary) 2d SCFT. The bootstrap approach is then used to obtain bounds on that theory, 
which can be directly translated to the four-dimensional theories. Supersymmetry plays a key role in that analysis, whereas we are interested in non-SUSY states in general. Also, the 2D CFTs obtained there are non-unitary, as is manifested in their central charge and Kac-Moody levels, whereas the theory that we will obtain is unitary. These two differences lead us to believe that the states that we are interested in are the ones that are lifted from being 2D in their case once $g_{ym}\not= 0$, in which case one needs to go to high energies (high Fermi level in our language) to have some measure of control. It will be interesting to explore whether the $W$ symmetry that they find there is part of the extended $W$-symmetry that exists in our suggested realization of "strange metals".
	
2. The extremal black holes dual to the fermi surface have zero entropy, and are therefore singular. Such black holes have been dubbed Extremal Vanishing Horizon (EVH) \cite{SheikhJabbaria:2011gc} black holes 
since the entropy, and thus the area of the horizon, shrinks to zero size along one of the dimensions for some subset of the solution space. Unlike the case of Kerr/CFT \cite{Guica:2008mu,Castro:2010fd}, EVH black holes have an $AdS_3$ near-horizon, and are thus proposed to be dual to fully (non-chiral) $2d$ CFTs. Due to the fact that the circle in the $AdS_3$ has vanishing periodicity at the horizon, it is known as a 'pinching' $AdS_3$. 

In \cite{SheikhJabbaria:2011gc} a conjecture has been made regarding the dual low-energy CFT for such black holes, which was further developed in \cite{deBoer:2011zt,Johnstone:2013eg,Johnstone:2013ioa}. The vanishing horizon area means that the background is singular, 
and the EVH/CFT proposal gives a prescription, which is different from ours, for regulating 
this singularity by rescaling $G_N$ to zero in order to obtain a finite entropy. 
The entropy can then be reproduced by the Cardy formula for a theory with central 
charge $c \propto N^2 \epsilon$ which must be kept fixed in the large-$N$ limit.
The case we consider, however, does not require us to rescale $G_N$, which would seem 
unnatural from the point of view of the duality between type $IIB$ SUGRA and ${\mathcal N} = 4$ SYM. 

%%%%%%%%%%%%%%%%%%%%%%%%%%%%%%%%%
\subsection{Summary of Results}
In this work we will be interested in finding the low energy effective field theory about a special class of states in ${\cal N}=4$ SYM theory. This class of states are Fermi surface operators built out of the partons of the fermionic $\su(1,1)$ sector of the ${\cal N}=4$ theory, denoted by $\rho^a_k$ where $a$ is an index of $su(N)$ and $k$ is the momenta of the parton (or the number of $\partial_{1{\dot 1}}$ derivatives acting on $\rho^a$ in the operator notation), i.e.,
\begin{equation}
\Pi_{a=1}^{N^2-1}\Pi_{k=0}^K \rho^a_k=\Pi_{a=1}^{N^2-1}\Pi_{k=0}^K \partial_{1{\dot 1}}^k\rho^a
\end{equation} Such operators are ground states in this sector,  for appropriate total charge and angular momenta, and they are exact eigenvector of the dilatation operator.  We will be interested in the limit of large $K$, and  $\epsilon=1/K$ will emerge as a new small expansion parameter, which will play a crucial role below. 

At tree level, excitation around the fermi surface are multiparticle states made out of particles above the fermi surface and holes below the fermi surface, with a global gauge invariance constraint, i.e, the states of  
\begin{equation}
\label{left}
{ SU(N)_N \otimes SU(N)_N \over \text{global } SU(N) }
\end{equation}  
We evaluate the corrections to the anomalous dimensions of such excitations to order $g^4$, and show that the latter theory splits into two sector - one receives no consequential anomalous dimension, in the large $K$ limit, and another which receives a positive anomalous dimension of order 1 times the appropriate power of $g$. The former, low energy sector, is that of a chiral sector of the gauged model 
\begin{equation}
\label{left}
{ SU(N)_N \otimes SU(N)_N \over  SU(N)_{2N} }
\end{equation}  

At order $g^2$ there is a simple expression for the anomalous dimension 
\begin{equation}
\dD_2 = \sum_k \rho^a_k \check \rho^a_k + {1 \over 2 N}  \sum_{u>0} {1 \over u} J^a_{-u} J^a_u 
\end{equation}
where $J^a_u$ is a Kac-Moody current with level $2N$ for small $u$ and large $K$ relevant for the low energy  excitations. This clearly shows that a gap (which we show to be $\Op(1)$) opens up between states which are annihilated by $J^a_u$ for $u>0$ which remain light, and those which are not. The light states are exactly those of  \eqref{left}. We show that this persists to order $g^4$, where the light states receive only corrections which are suppressed by $1/K$ whereas "heavy states" receive corrections which are $\Op(1)$ (times $g^4$). If this persists to strong coupling we can expect that the states of  \eqref{left} remain light there, whereas the "heavy" states receive arbitrarily large anomalous dimensions. 

\subsection{Outline of the paper}

The outline of this paper is as follows. Sections 2,3 provide some background material. Section 2.1 is a quick introduction to the "strange metal" CFT. Section 2.2 provides, for completeness, a discussion of the dual gravity configuration. Section 3 sets up the computations that we will do later, by detailing the $\psu(1,1|2)$ and fermionic $\su(1,1)$ sectors and by introducing the Fermi surface state around which we will expand. Section 4 computes the order $g^2$ anomalous dimension and the origin of the $SU(N)_{2N}$ gauging. Section 5 carries out the same computation in the limit of large angular momentum, or large Fermi energy of the Fermi surface. In this limit a new diagrammatic scheme emerges, which vastly simplifies the computation of the anomalous dimension. In Section 6 we check our conjecture at two loop level using the techniques developed in Section 5. In Section 7, we discuss our results and point out future directions. 

%%%%%%%%%%%%%%%%%%%%%%%%%%%%%%%%%%%%%%%%%%%%%%%%%%%%%%

\section{"Strange metals", Fermi surfaces in ${\cal N}=4$ and degenerate Black holes in $AdS_5 \times S^5$}\label{blackhole}

\subsection{Strange Metals in 1+1 dimensions}

 In this section, we will review the "strange metal" coset models in 1+1 dimensions, following \cite{Gopakumar}. Consider a $SU(N)$ gauge theory in 1+1 dimensions coupled minimally to adjoint fermions. The Lagrangian is
\begin{equation}
{\cal{L}} = \tr \left[ \bar \Psi (i \gamma^\mu D_\mu \Psi - m- \mu \gamma^0) \Psi \right] - {1 \over  2 g_{YM}^2} \tr F^2
\end{equation}
In the high density limit $\mu \gg m, g_{YM} \sqrt N$, the ground state is just a fermi surface. The low energy excitations around this state are Dirac fermions interacting with each other via gauge fields. The effective Lagrangian relevant at  low energies is
\begin{eqnarray}\nonumber
{\cal {L}}_{eff}   &=& \tr \left[ \psi_R^\dagger (\partial_\tau - \partial_x)\psi_R + (A_\tau + A_x)  [ \psi_R^\dagger , \psi_R] \right] + 
\nonumber \\ &&  \hspace{10mm} \tr \left[\psi_L^\dagger (\partial_\tau + \partial_x)\psi_L + (A_\tau -A_x)  [ \psi_L^\dagger , \psi_L]\right]  -
  {1 \over  2 g_{YM}^2} \tr F^2
\end{eqnarray}
where the fermions $\psi_{L(R)}$ are left(right) moving fermions defined  from the microscopic $\Psi$  fermions by linearizing around the fermi surface. To see the  emergence of coset more clearly, it is useful to trade the Dirac fermions for a pair of Majorana fermions 
\begin{equation}
\psi_{L,R}  =  {1 \over \sqrt 2} ( \psi^1_{L,R} + i \psi^2_{L,R} )
\end{equation}
Now the Lagrangian becomes
\begin{eqnarray}
{\cal {L}}_{eff} &&= -  {1 \over  2 g_{YM}^2} \tr F^2 \\ 
&& + {1 \over 2} \tr \left[ \psi^a_R (\partial_\tau - \partial_x)\psi^a_R + (A_\tau + A_x)   \psi^a_R  \psi^a_R + \psi_L^a (\partial_\tau + \partial_x)\psi^a_L + (A_\tau -A_x) \psi^a_L  \psi^a_L\right] \nonumber
\end{eqnarray}
In the strong coupling limit ($ g_{YM} \rightarrow \infty$) the gauge fields decouple to give just free fermions with constraints that the currents $J_R = \psi^a_R \psi^a_R$ and $J_L = \psi^a_L \psi^a_L$ vanish.  Each Majorana fermion is equivalent to a $SU(N)_N$ WZW model. Also the currents $J_L,J_R$ that must vanish obey a $SU(N)_{2N}$ Kac-Moody algebra.  Hence the low energy theory is a $CFT_2$ based on coset
\begin{equation}
\frac{SU(N)_N \otimes SU(N)_N}{SU(N)_{2N}}
\end{equation}
The above CFT has ${\cal N} = (2,2)$ supersymmetry as shown in \cite{kent}. The central charge of the theory is 
\begin{equation}
c = {N^2-1 \over 3}
\end{equation}
For the $N=2,3$ case, the full conformal primary operator spectrum was analyzed in \cite{Gopakumar} and the operators are constructed explicitly from the fermions. In addition, a partial list of operators for $N \ge 4$ is given (for $N=4,5$, the chiral, in the sense of SUSY, primary content of the theory has been worked out in \cite{Schomerus} using group theoretic techniques). 

In this work, we will be interested in the chiral sector of this theory, i.e., all operators with conformal dimension $(0,h)$.

%\subsection{Near horizons of rotating black holes and chiral field theories} 
%The fact that the fermi surface state that we describe in section \ref{Neq4FS} is dual to the (fast) rotating black holes discussed in section \ref{blackhole}, has consequences for the low energy theory on both sides. The low energy theory of the black holes is described just by restricting oneself to the near horizon limit. On the other hand, in this work we find that the low energy excitations about the fermi surface is a chiral sector of coset CFTs described above. This suggests that the chiral  CFT must be dual to the near horizon limit of the rotating black hole. We comment on various aspects of this below

The coset CFT described above is a special case of the class of CFTs
\begin{equation}
\frac{SU(N)_k \otimes SU(N)_l}{SU(N)_{k+l}}
\end{equation}
having ${\cal N}=(1,1)$ supersymmetry. The case with $k=N,l=1$ has been well studied. It is known to have an extended $W_N$ symmetry i.e one chiral current for each spin $s=2,3..N$ (see for example \cite{Bouwknegt}). These are vector like models with central charge $c \sim N$ in large N limit, which have been proposed \cite{Gaberdiel:2010} (see \cite{Gaberdiel:2012} for a review) to be dual to with Vasiliev higher spin theories on AdS \cite{Vasiliev3} in the bulk. 

In the case we are interested in with $k=l=N$, there are many additional chiral currents apart from the $W_N$ currents. These form a much larger higher spin algebra whose consequences have not yet been worked out fully. They are "matrix like" models with central charge $c \sim N^2$ in large N limit.  In fact, there is a hagedorn growth in the number of higher spin currents suggesting that the bulk dual must have a much bigger gauge symmetry than Vasiliev theories, maybe even full string theory. 

%Our work then suggests a duality between the chiral sector of the above CFT's and string theory at the near horizon geometry of the rotating black hole described above. The hagedorn growth in the CFT arises due to the fact that the near horizon contains singularities which are resolved by including the full string spectrum.

% \textbf{New phase of GR at singular BH:}

% \textbf{Not Kerr/CFT:}

\subsection{Black hole and Fermi surfaces}

Most of this work deals with the dynamics of excitations in the fermionic $\su(1,1)$ sector, and to a lesser extent in the $\psu(1,1|2)$ sector \cite{Beisert}, in the weak coupling limit. Our motivation, however, is also in understanding a class black holes in $AdS_5\times S^5$, which are their conjectured duals \cite{Dori}. For completeness, we provide of brief discussion of these black holes. 

The specific black holes are describe within 
%In , a specific Supergravity solution with small amount of supersymmetry was studied, and was %proposed to be dual 
%to a fermi-surface configuration in the $PSU(1,1|2)$ sector of ${\cal N}=4$ SYM , which we will 
%eview in more detail later. 
the consistent truncation of type $IIB$ Supergravity on $AdS_5 \times S^5$ described in ref. \cite{cvetic}.
The field content consists of the metric, two neutral scalars and three abelian $U(1)$ fields.
The bosonic part of the supergravity action is 
\begin{align}
\label{lagrangian}
S = &\int d^5 x \sqrt{-g} \left[ R - \frac{1}{2} \sum_{\alpha=1}^{2}{(\partial \varphi_{\alpha})^2} + \sum_{i=1}^{3}{\left(4 l^{-2} X_i^{-1} - \frac{1}{4} X_i^{-2} \mathcal{F}^i_{\mu \nu} \mathcal{F}^{i \mu \nu}\right)}\right] + \nonumber \\
&\int d^5 x \frac{1}{24} |\epsilon_{ijk}| \epsilon^{uv \rho \sigma \lambda} \mathcal{F}^i_{uv} \mathcal{F}^j_{\rho \sigma} A_{\lambda}^k.
\end{align}
Here,  $l_{AdS} \equiv l$ is the AdS radius, $A^i$ are the three $U(1)$ gauge fields, and
$X_i$ are three uncharged scalars, constrained by $X_1 X_2 X_3 = 1$ and parameterized by
$X_1 = e^{-\frac{1}{\sqrt{6}}\phi_1 -\frac{1}{\sqrt{2}} \phi_2}, X_2 = e^{-\frac{1}{\sqrt{6}}\phi_1 +\frac{1}{\sqrt{2}} \phi_2}, X_3 = e^{\frac{2}{\sqrt{6}}\phi_1}.$
%\end{align}
\
Black hole solution in this SUGRA model were found in \cite{Chong:2005da} and generalized in \cite{Mei2007}. 
We follow the latter's conventions. 
The most general known class of solutions, describing a black hole, is parameterized by 5 numbers $(E,J_\psi,J_\phi,Q_1=Q_2,Q_3)$ - the mass, angular momenta along the two independent 2-planes, and the 3 $U(1)$ charges. In our convention
\begin{align}
\label{angular momentum relation}
J_{\phi}=J_L + J_R,\ \ \ J_{\psi}=J_L - J_R.\ \ \ Q_i=\frac{\hat{Q}_i}{l}.
\end{align}
where the ${\hat Q}$ are the field theory charges normalized to be dimensionless. 
%The relation between these charges and the field-theory charges ${\ca is
The charges $J_{\phi}$, $J_{\psi}$, $Q_1(=Q_2)$, $Q_3$ and $E$ are then further parameterized, as in \cite{Mei2007}, by $\delta_1$, $\delta_3$, $m$, $a$, $b$, which are the parameters that appear in the SUGRA background. Our $l$ is denoted there by $1/g$.

We will focus on a subset of these solutions given by
$Q_3 = 0$ and $J_L-J_R = J_\psi = 0$, which is equivalent to setting $b = 0$ and $\delta_3 = 0$.
The metric is given by
\begin{equation}
\begin{split}	
\label{metric2}
ds^2 = H_1^{2/3} \Biggl\{ &  \left(x^2 + y^2\right) \left(\frac{dx^2}{X} + \frac{dy^2}{Y} \right) -
			\frac{ X \left( dt - y^2 d \sigma \right)^2}{\left(x^2 + y^2\right) H_1^2} + \Biggr. \\
			\Biggl.& \frac{Y \left[dt + \left(x^2 + 2 m s_1^2\right) d\sigma \right]^2}{\left(x^2 + y^2\right) H_1^2}  + y^2 x^2 d\chi^2 \Biggr\}
\end{split}
\end{equation}
where we define $c_1 = cosh(\delta_1)$,$s_1 = sinh(\delta_1)$ and $\Sigma_a = 1-a^2/l^2$.
The functions used in the metric are
\begin{equation}
\begin{split}
X &= -2m + \left(a^2 + x^2\right) + l^{-2}\left(a^2 + 2m s_1^2 +x^2\right) \left( 2m s_1^2 + x^2 \right) \\
Y &= \left(a^2 - y^2\right) \left( 1 - l^{-2} y^2\right)  \\
H_1 &= 1 + \frac{2 m s_1^2}{x^2+y^2}.
\end{split}
\end{equation}
In addition, the gauge field and scalar backgrounds are
\begin{equation}
\begin{split}
A^1 &= A^2 = \frac{2 m s_1 c_1 (dt - y^2 d\sigma)}{(x^2 + y^2)H_1} \\
A^3 &= \frac{2 m s_1^2 y^2 d\chi}{(x^2+y^2)}  \\
X_1 &= X_2 = H_1^{-1/3} \\
X_3 &= H_1^{2/3}.
\end{split}
\end{equation}

Finally, using the AdS/CFT relation 
\begin{equation}
\frac{\pi l_{AdS}^3}{4 G_5} = \frac{N^2}{2}.
\end{equation}
the black hole's global charges can be written as
\begin{equation}
\begin{split}
\label{BH_charges}
J_{\phi} &= \frac{\pi}{4 G_5} \frac{2 m a \left(1+ s_1^2\right)} {{\Sigma_{a}}^2} =N^2 l^{-3} \frac{ m a \left(1+ s_1^2\right)} {{\Sigma_{a}}^2} \\
Q_{1} &= Q_{2} \equiv Q = \frac{\pi}{4 G_5} \frac{2 m s_{1} c_{1}}{\Sigma_{a}}=N^2 l^{-3} \frac{ m s_{1} c_{1}}{ \Sigma_{a}} \\
E &= \frac{\pi}{4 G_5} \frac{m [ (2(l^{-4} a^4 + \Sigma_a + 1) + l^{-2} a^2 (\Sigma_a -2))s_1^2 + \Sigma_a  + 2] }{\Sigma_a^2} 
\end{split}
\end{equation}

As explained in \cite{Micha}, the event horizon of the black hole is obtained by
changing to the asymptotically $AdS$ coordinates 
\begin{equation}
\begin{split}
\label{x_to_rho}
& x^2 =r^2-\frac{4}{3} m s_1^2 \\
& y^2=a^2 cos^2\theta
\end{split}
\end{equation}
and solving $\Delta(r) \equiv X(x) x^2 = 0$.
The extremal limit occurs when $X(x) \propto x^2$.
The (degenerate) horizon is then at $x = 0$. 
For convenience we changed variables to $x^2 = z$. Now, $X(x) = (z-z_1)(z-z_2)$ and $z_2$ serves as the off-extremality parameter. 

In the limit of large angular momentum, i.e. $a \rightarrow l$, with all other parameters fixed, we have from (\ref{BH_charges}) that
\begin{align}
J_{\phi} &\propto N^2 \frac{1}{\left(1-\frac{a}{l}\right)^2} \\ 
Q_1 = Q_2 &\propto N^2 \frac{1}{1-\frac{a}{l}}.
\end{align}
This scaling suggests defining the (finite) ratio
\begin{equation}
\label{alpha_scaling}
\alpha = \frac{J_{\phi} / N^2}{(Q_1 / N^2)^2}.
\end{equation}
At the value $\alpha=2$ the extremal black hole satisfies a $1/8^{th}$ BPS bound 
\begin{equation}
E=J_{\phi}+J_{\psi}+Q_1+Q_2+Q_3=J_{\phi}+2Q_1
\end{equation}
and has zero entropy. This can be seen by expressing the remaining parameters in terms of $\alpha$ and $z_2$, with $l=1$ for simplicity 
\begin{equation}
\begin{split}
m &= \frac{\left(a^2+z_2\right) \left(1+z_2\right)}{2}+\frac{2 a^2}{\alpha^2}+\frac{a^3+2 a z_2}{\alpha} \\
m s_1^2 &= \frac{a}{\alpha} \\
z_1 &=  -\frac{4 a+\left(a^2+1+z_2\right) \alpha }{\alpha},
\end{split}
\end{equation}
where now
\begin{equation}
\begin{split}
S &= N^2 \frac{2\pi \sqrt{z_2} \left(a^2+\frac{2 a}{\alpha }+z_2\right)}{\Sigma_a} \\
T &= \frac{\sqrt{z_2} \left(4 a +\alpha + a^2 \alpha +2 \alpha z_2\right)}{4 a \pi +2 \pi \alpha   \left(a^2+z_2\right)}.
\end{split}
\end{equation}
Clearly $S=T=0$ at $z_2=0$. 

Massless scalar perturbations in this black hole background have been studied 
and exhibit the spectrum of a free fermion bilinear in a $1+1$ CFT \cite{Micha}, i.e., a chiral current, which is expected to appear in the spectrum of the Strange Metal \cite{Gopakumar}.

The scaling expressed by a fixed $\alpha$ in  (\ref{alpha_scaling})
is similar to that of a one-dimensional fermi surface.
Suppose we build a fermi surface of a single free fermion living 
on a circle with radius $1$, so that the momentum is discretized.
When building a fermi surface, a fermion at energy level $n$ contributes a single unit to 
the total charge $Q$ and $n$ units to the total angular momentum $J$. 
Therefore, a fermi surface with $K \gg 1$ states has 
$Q=K$ and $J \approx K^2$, so that $J/Q^2 \approx 1$.  
If the fermion is in the adjoint of $SU(N)$, 
there are $N^2 - 1$ states at every level which one needs to take into account. 
In the following section we show how such a fermion can be embedded in $\mathcal N = 4 SYM$. 

\section{Constructing a $1D$ fermi Surface in $\N=4$ Super Yang-Mills Theory}

In this section we show how the fermi surface-like scaling of the black hole can be realized in $\N=4$ SYM. We introduce the $\psu(1,1|2)$ and fermionic $\su(1,1)$ sectors, and describe their ground states, which are conjectured to be dual to the $\alpha=2$ black holes, for large angular momenta. 

An explicit computation of the anomalous dimension for these fermi surface ground states, carried out in \cite{Dori}, has shown that it is weakly renormalized at two loops, i.e., the corrections to the conformal dimensions is suppressed by powers of $Q/N^2$ relative to the classical dimension. Here we study in greater detail excitations around this fermi surface, finding  evidence that the low-energy excitations about this fermi surface behave like a (chiral) Strange Metal in $1+1$ dimensions.

In section \ref{Neq4not} we review the ${\cal N}=4$ multiplet and symmetry generators  to set up the notation. In section \ref{Neq4sectors} and section \ref{Neq4psu} we discuss the closed subsectors of SYM theory paying close attention to $\psu(1,1|2)$ and fermionic $\su(1,1)$ sectors. In section \ref{Neq4FS} we discuss the fermi surface ground states of these sectors. 

%%%%%%%%%%%%%%%%%%%%%

\subsection{$\N = 4$ SYM Notations}\label{Neq4not}
Our conventions for the $\N=4$ multiplet are\footnote{Throughout this paper we follow the notation of \cite{Beisert}.}: (1) the gauge field-strength $F_{\alpha\beta}$ and $\bar F_{\dot\alpha\dot\beta}$, (2) the gauginos $\Psi_{\alpha a}$ and $\bar\Psi_{\dot\alpha}^a$ and (3) The complex scalars $\Phi_{ab}$ with the antisymmetry $\Phi_{ab}=-\Phi_{ba}$ (and $ (\Phi_{ab})^\dag  = \bar \Phi^{ab} = \frac12\epsilon^{\,abcd}\Phi_{cd}$). 
The undotted Greek letters ($\alpha,\beta,\ldots$), dotted Greek
letters ($\dot\alpha,\dot\beta,\ldots$) and Latin letters
($a,b\ldots$) stands for $\SU(2)_L$, $\SU(2)_R$ and $\SU(4)$
fundamental indices, respectively. Raising and lowering the $\SU(4)$ indices
changes between the fundamental and anti-fundamental
representations.

The gauge group is $G=\SU(N)$, and all fields transform in the
adjoint representation. When we will need to be specific about the
gauge group structure we will write all fields as $\W=\W^at^a$ with
$a=1,\ldots \dim G$, and $t^a$ are generators of $\SU(N)$
\footnote{We use the same letters for gauge group and $\SU(4)$
indices, it will be clear to distinguish between them from the
context. After this section only gauge group indices are used.}. 
The covariant derivative is
\begin{equation}
    D_{\alpha\dot\alpha}\W =(\sigma^\mu)_{\alpha\dot\alpha}\left(\pd_\mu\W-i[A_\mu,\W]\right)~,
\end{equation}
where $\mathcal W$ are the partons in the theory $\mathcal W \in
\begin{Bmatrix} D^{k}F, & D^{k}\Psi_i, & D^{k}\Phi_{ij}, &
D^{k}\bar\Psi^i, & D^{k}\bar F
\end{Bmatrix}$.%, with all Lorentz indices symmetrized. 
We use, as in \cite{Beisert}, $\check\W^A$ for functional derivatives
with respect to the partons
\begin{equation}
    (\check\W)^a = \fdf{}{(\W)}\spc
    a=1,2\ldots\dim G~. 
\end{equation}

The generators of the $\psu(2,2|4)$ algebra are:
\begin{itemize}
   \item The compact bosonic $\su(2)_L\times\su(2)_R\times\su(4)$ generators ${\mathfrak{L}^\alpha}_\beta$ , $\mathfrak{\bar L}^{\dot\alpha}_{~\dot\beta}$ , ${\R^a}_b$.
   \item The non-compact bosonic translation, dilatation and special conformal generators $\mathfrak{P}_{\alpha\dot\alpha}$, $\mathfrak{D}$, $\K^{\dot\alpha\alpha}$.
   \item The supercharges $\mathfrak{Q}^a_{\alpha}$, $\mathfrak{\bar Q}_{\dot\alpha a}$ and super-conformal supercharges $\mathfrak{S}_a^{\alpha}$, $\mathfrak{\bar S}^{\dot\alpha a}$~.
\end{itemize}

%%%%%%%%%%%%%%%%%%%

\subsection{Closed Sectors in $\N =4$ SYM} \label{Neq4sectors}
Below we will expand the anomalous dimension operator $\dD=\D-\D_0$, as well as a subset of the other operators in $\psu(2,2|4)$ algebra, in an expansion in $g^2 = \frac{g_{ym}^2N}{8\pi^2}$. I.e., 
\begin{equation}
    \dD = \sum_{n=2}^\infty\dD_n g^{n}~.
\end{equation}
We will choose a regularization scheme such that
operator mixing occurs only between operators with the same
zero-coupling dimension,
and where the
Poincar\'e group and R-symmetry do not receive quantum corrections.

We will focus below on a specific sector of the theory, i.e., a set of states closed under operator  mixing, in this scheme. Such sectors have been classified in \cite{Beisert}. 
Since $\N=4$ is a rich theory with complicated dynamics, the sectors offer a significant simplification, as they allow one to isolate and study the dynamics of a smaller subset of partons. 
Some sectors, such as the $\su(2)$ sector, in fact contain a finite number of partons. However, these are too restricting for our purpose as they do not contain the large number of fermions needed to construct a fermi surface. 

%%%%%%%%%%%%%%%%%%%%%%%%%%

\subsection{The $\psu(1,1|2)$ Sector and its fermionic Subsector}\label{Neq4psu}

The $\psu(1,1|2)$ sector, however, is much richer. This sector has been studied extensively in the literature \cite{Thesis,Zwiebel,Beisert2007}, and it is obtained by demanding the following relations between the charges
\begin{equation}
\label{psu_constraint}
\Delta_0 = 2 J_L + \hat{Q}_1 + \hat{Q}_2 + \hat{Q}_3 = 2J_R + \hat{Q}_1 + \hat{Q}_2 - \hat{Q}_3
\end{equation}
where $\Delta_0$ is the classical scaling dimension, $J_L$ and $J_R$ are the $SU(2) \times SU(2)$ quantum numbers, and
$\hat{Q}_1$, $\hat{Q}_2$ and $\hat{Q}_3$ are the $SU(4)$ $R$ charges spanning the Cartan subalgebra of $SO(6)\cong SU(4)$.
The above relations allow only states built out of four types of partons
\begin{align}
\phi^1_{k} &\equiv \frac{1}{(k)!} D^k_{1\dot{1}} \phi_{24} & \phi^2_{k} \equiv \frac{1}{(k)!} D^k_{1\dot{1}} \phi_{34} \nonumber \\
\psi_{k} &\equiv \frac{1}{(k+1)!} D^k_{1\dot{1}} \psi_{14} & \bar{\psi}_{k} \equiv \frac{1}{(k+1)!} D^k_{1\dot{1}} \bar{\psi}^1_1.
\end{align}
Here, $D_{1\dot{1}}$ is the covariant derivative $D_{\alpha \dot{\alpha}}$ with $\alpha = \dot{\alpha} = 1$.

The symmetry inherited by this sector from the full theory is a 
$\psu(1,1|2) \times \psu(1|1)^2$  algebra. It includes 
\begin{itemize}
\item The $\mathfrak{su}(1,1|2)$ symmetry, generated by 
\begin{align}
    J^0=&\, - \LL+2\D_0+\dD &
    R^0 =&\, \R^2_{\ 2}-\R^3_{\ 3} \\
    J^{++} =&\, \mathfrak{P}_{1 \dot 1} &
    J^{--} =&\, \K^{1 \dot 1}
    \\
    R^{22} =&\, \R^3_{\ 2}&
    R^{33} =&\, \R^2_{\ 3}
    \\
    Q^{+i}=&\,\mathfrak{Q}^i_1&
    \bar{Q}^{+i}=&\,\mathfrak{ \bar{Q}}_{\dot 1i}& 
    \\
    Q^{-i}=&\,\mathfrak{ \bar{S}} ^{\dot 1i}&
    \bar{Q}^{-i}=&\,\mathfrak{S}^1_{i}&
\end{align}
where $i=2,3$, and $\LL$ is the length (i.e. parton number) operator. 
\item The $\mathfrak{psu}(1|1)^2$ symmetry generated by
\begin{align}
    I^+=&\,\mathfrak{ \bar{Q}}_{\dot 2 4}&
    I^-=&\,\mathfrak{S}^{2}_{1}
    \\
    \bar I^+=&\,\mathfrak{Q}^{1}_{2}&
    \bar I^-=&\,\mathfrak{ \bar S}^{\dot 24}
    \\
    \dD&&
    \LL
\end{align}
With the relation
\begin{equation}
    \dD=2\left\{I^+,\bar I^-\right\}=2\left\{I^-,\bar I^+\right\}~.
\end{equation}
\item In addition, as shown in \cite{Beisert2007}, there is also an $SU(2)$ automorphism which exists only within the sector, under which both $\phi^1_k$ and
$\phi^2_{k}$ are singlets for all $k$, while $\psi_k$ and $\bar{\psi}_k$ are a doublet. We will refer to is as the custodial $SU(2)_c$ symmetry.
\end{itemize}

When constructing a fermi-surface operator within this sector, 
one may still worry that the scalars cause instabilities and produce large mixing effects. 
It is possible to restrict to a further closed subsector, namely the fermionic $\su(1,1)$ sector discussed in \cite{Beisert2007} (there it is called the fermionic $\ssl(2)$ sector). This sector consists of the set 
\begin{equation}
\left\{\psi_k^a=\frac{1}{(k+1)!} D^k_{1\dot{1}} \psi_{14},\ k=1..\infty,\ a=1...dim(G)\right\}.
\end{equation} That this is a closed sector can shown directly using the oscillator formalism \cite{Beisert2007} or by using the $SU(2)$ automorphism above. 

%%%%%%%%%%%%%%%%%%%%%

\subsection{The fermi Surface ground state}\label{Neq4FS}
\label{sec-gs}
The simplest fermi surface is constructed using only a single fermion.
It contains derivatives of the fermion 
ranging from $0$ to some large $K$, and is given by (with explicit gauge indices)
\begin{equation}\label{fermi-sea-1dim}
    \Op^{(K)}
    =\prod_{n=0}^{K}\prod_{n=0}^{dim(G)}\psi_{n}^a~. 
\end{equation}
All fermionic operator are evaluated at the same space point, correspondingly the
expression can be viewed as a state in radial quantization.
This operator was studied in ref. \cite{Dori,Micha}.

The 1-dim fermi-surface at zero coupling, large $N$ and large $K$ has
the following charge, dimension and angular momenta
\footnote{details about the definitions of the charges are found in Appendix A of \cite{Dori}}:
\begin{subequations}
\begin{align}
\label{1dimch}
    (\hat Q_1,\hat Q_2,\hat Q_3)=&\,{1 \over 2}\sum_{n=0}^{K}\sum_{a=1}^{\dim G}\left(1,1,1\right) 
    \approx\,\left({N^2 K \over 2},{N^2 K \over 2},{N^2 K \over 2}\right),
\\
\label{1dimdim}
    \Delta_0 =\,&\sum_{n=0}^{K}\sum_{a=1}^{\dim G}\left(\frac32+n\right)=(N^2-1)\frac{(K+3)(K+1)}{2}
    \approx\,\frac{N^2K^2}{2},
\\
\label{1dimang}
    \left(J_L,J_R\right)=&\,\sum_{n=0}^{K}\sum_{a=1}^{\dim G}\left(\frac{n+1}{2},\frac{n}{2}\right)=(N^2-1)\frac{K(K+1)}{4}\left(\frac{K+2}{K},1\right)\cr
    \approx&\,\left(\frac{N^2K^2}{4},\frac{N^2K^2}{4}\right)~. \end{align}
\end{subequations}

This operator is the unique ground state in the fermionic sector, with these charges (or chemical potential). It does not mix with any other operators in the theory and it is thus an eigenstate of the dilatation operator \cite{Dori}.
Furthermore, it was found in ref. \cite{Dori} that the dimension of this operator, 
to order $\mathcal{O}(g^4)$, is the classical dimension with corrections of 
order of the inverse of the (large) charge. Explicitly, the computation yields
\begin{align}
\label{g2dil}
    \D\ket{\Op^{(K)}}     =
    &(N^2-1)\frac{(K+3)(K+1)}{2}\left[1+\frac{4\,\left(g^2-g^4\right)}{K+3} + O(g^6)
    \right]\ket{\Op^{(K)}}~. \end{align}
It was conjectured there that this $O(1/K)$ suppression survives the strong coupling limit, so that the fermi surface has finite anomalous dimensions to all orders in $g$.  

However, the charges of these states do not match the charges of black holes in section \ref{blackhole}. To find states which do have the charges of black holes we can rotate by the custodial $SU(2)_c$. More precisely, the Fermi surface that we have just constructed is the maximal $J_{c,3}$ vector in a custodial $SU(2)_c$ representation with "spin" $(N^2-1)\times(K+1)/2$. We can rotate this state, within the same representation, such that it has
$\langle J_{c,3} \rangle=0$. Very qualitatively (assume $K$ to be even for simplicity) we can think about the state as  
\begin{equation}
\label{operator}
\mathcal{O}^{(K)} = Sym\left[\prod_{a,b=1}^{dim(G)}\prod_{j=0}^{{K \over 2}-1} \psi_j^a \prod_{m={K\over 2}}^{K-1} \bar{\psi}_m^b \right]
\end{equation}
where $Sym[\text{  }]$ stands for a symmetrization of the operator with respect to the $\psi, \bar{\psi}$ fermions,
placing the operator in the highest $SU(2)$ state, with $J^2 = K N^2(K N^2+1)$ and $J_z=0$. 
In any case, however, we will use the description in \eqref{fermi-sea-1dim} since the two descriptions are equivalent under $SU(2)_C$. 

One can easily compute the charges for such an operator
\begin{align}
\label{charges_eq}
\hat{Q}_1 &= \hat{Q}_2 \equiv \hat{Q} = {N^2 K \over 2} \nonumber \\
\hat{Q}_3 &= 0 \nonumber \\
J_L &= N^2 \left(\sum_{j=0}^{{K\over 2}-1}\frac{j+1}{2} + \sum_{m={K\over 2}}^{K-1}\frac{m}{2}\right) = {N^2 K^2 \over 4} \nonumber \\
J_R &= N^2 \left(\sum_{j=0}^{{K\over 2}-1}\frac{j}{2} + \sum_{m={K\over 2}}^{ K-1}\frac{m+1}{2}\right) = {N^2 K^2 \over 4} = J_L \nonumber \\
\Delta_0 &= 2J_R + \hat{Q}_1 + \hat{Q}_2 = {N^2 \over 2}(K^2 +2 K)
\end{align}
with $\Delta_0$ the classical scaling dimension. These charges match the black hole charges given in section \ref{blackhole}. 

Although this operator looks quite different from the simpler fermi surface operator presented in (\ref{fermi-sea-1dim}), the two are related by a simple rotation in the $SU(2)$ automorphism. 
Hence in all computation in field theory within the sector, the two give the same answer. In particular,
we expect the corrections to dimensions are suppressed in this case too. We also expect same low energy excitations about the fermi surface  in the two cases. For simplicity, we will use the simple fermi surface operator given in (\ref{fermi-sea-1dim}) for all subsequent computations.

%%%%%%%%%%%%%%%%%%%%%%%%%%%%%%%%%%%%%%%%%%%%%%%%%%%%%%%%%%%%%%%
\section{Emergence of a chiral "strange metal"}\label{DiscreteD2}
%%%%%%%%%%%%%%%%%%%%%%%%%%%%%%%%%%%%%%%%%%%%%%%%%%%%%%%%%%%%%%%

%%%%%%%%%%%%%%%%%%%%%%%%%%%%%%%%%%%%%%%%%%%%%%%%%%%%%%%%%%%%%%%
\subsection{Excitations of the fermi surface in free field theory} 
Let us first comment on the free theory, i.e the theory with $g=0$. Let the Hilbert space of small excitations around the 1+1 dimensional fermi surface be ${\cal H}_F$. We will give a more precise definition of ${\cal H}_F$ later.  Since each complex fermion can be written in terms of two majorana fields, the states in ${\cal H}_F$ are governed by a chiral $SU(N)_N \otimes SU(N)_N$ WZW model (both at the level of enumeration of states, and at the level of their energies). Since only SU(N) gauge singlet excitations are allowed, there is a global $SU(N)$ constraint. Thus the free theory is described by \footnote{A way to implement the global constraint is to introduce an $SU(N)$ orbifold in target space, and keep only those states which are uncharged under this $SU(N)$, see for example \cite{Gaberdiel:2014vca}. We thank M.~Gaberdiel for an interesting discussion on this point.}
\begin{equation}
{ SU(N)_N \otimes SU(N)_N \over \text{Global SU(N)}}.
\end{equation}

%%%%%%%%%%%%%%%%%%%%%%%%%%%%%%%%%%%%%%%%%%%%%%%%%%%%%%%%%%%%%%%
\subsection{1-loop dilatation operator}

As a first step it is convenient to define new operators by 
\begin{eqnarray}
\label{fermion_norm}
\rho^a_k = \sqrt{k+1} \psi^a_k \hspace{20mm} \check \rho^a_k = {\check \psi^a_k \over \sqrt{k+1}}
\end{eqnarray}
which satisfy  $(\rho^a_k)^\dagger = \check \rho^a_k$ and $\{ \rho^a_k , \check \rho^b_q\} = \delta_{k,q} \delta^{ab}$.

As mentioned before, the dilatation operator appears as the central extension of the $\psu(1,1)^2$ algebra and can be written as $\dD = 2 \{I^+ , \bar I^-\}$. 
In particular, 
\begin{equation}
\dD_2 = 2 \{ I_1^+ , \bar I_1^-\}.
\end{equation}
where generally $I=g\times I_1+\Op(g^3)$, and, as in \cite{Zwiebel} and in (3.17) of \cite{Dori}, 
\begin{eqnarray} \label{Iporig}
I_1^+ &=&   {1 \over \sqrt{2}} \sum_{k,q} \sqrt{k+q+2 \over (k+1)(q+1)} \tr \cno{\rho_k \rho_q \check \rho_{k+q+1}} \hspace{20mm}\\ 
\label{Imorig}
\bar I_1^- &=&   {1 \over \sqrt{2} N} \sum_{m,n} {\sqrt{n+ m + 2} \over \sqrt{(n+1) (m+1)}} \tr \cno{\rho_{n+m+1} \check \rho_n \check \rho_m} 
\end{eqnarray}
The one-loop dilatation operator $\dD_2$ can be written as \footnote{Our conventions of generators are $[t^a,t^b] = i f^{abc} t^c$,  $\tr_{\text{Fund}}(t^a t^b)= \delta^{ab}$,  $\tr_{Adj}(t^a t^b)=2 N \delta^{ab}$. This results in $f^{abc}f^{abd} = \delta^{cd} 2 N$  } 
\begin{eqnarray}\label{D2inrho}
\dD_2 =&  2 \smashoperator[r]{\sum_{k=0}^\infty} \rho^a_k \check \rho^a_k + {1 \over N} \smashoperator[r]{\sum_{\substack{q,m=0 \\ u=1}}^{\infty}} {1 \over u} \sqrt{q+1 \over q+ u +1} \sqrt{m+1 \over m+ u +1} \times \\ \nonumber
& [ i f^{eab} \rho^a_{m+u} \check \rho^b_m] [i f^{ecd} \rho^c_q \check \rho^d_{q+u}].
\end{eqnarray}
The derivation can be found in Appendix \ref{current_appendix}. 
It is now convenient to group the following combination of $\rho, \check \rho$ as
\begin{equation}\label{defJ}
{ J}^a_{
n} =- i f^a_{bc} \sum_{m=0}^{\infty}
\begin{cases}
\sqrt{m+1 \over m+n+1 }  \rho^b_{m} \check{\rho^c}_{m+n} &\mbox{if } n > 0 \\
\sqrt{m+1 \over m+|n|+1 }   \rho^b_{m+|n|} \check{\rho^c}_{ m} & \mbox{if } n < 0.
\end{cases}
\end{equation}
which satisfies $J^a_n \ket{\Op^{(K)}} = 0$ and  $(J^a_n)^\dagger = J^a_{-n}$, for all $n \ge 0$. Notice that $J$ satisfies (see Appendix \ref{current_algebra_appendix} for details) the following commutation relations
\begin{equation}\label{kacmoody}
[J^a_{m}, J^b_{n}] = i f^{abc} J^{c}_{(m+n)} + 2N\ m \delta^{ab} \delta_{m+n,0}+Residue
\end{equation}
where the Residue vanishes for $mn>0$, whereas for $m n < 0$ it is of order $\Op(1/K)$ when acting on small fluctuations of the Fermi surface provided $m,n \ll K$.
%\footnote{The normalized fermionic excitations around the fermi surface are constructed as
%\begin{equation}
%\Psi_i^\dagger = \sum_{p=1}^{K} f^i_{a,p} \ \check \rho^a_p  + \sum_{p=K+1}^\infty  f^i_{a,p} \ \rho^a_p
%\end{equation}
%where $f^i_{a,p}$ are some orthonormal set satisfying $\sum_{a,p=1}^\infty {f^i}^*_{a,p} f^j_{a,p} = \delta^{ij}$. States with excitations "close" to fermi surface have all the excitations within the interval $[K-s,K+s]$ in large $K$, fixed $s$ limit, i.e $f^i_{a,p}=0$ for $p \notin [K-s,K+s] $ See Appendix \ref{current_algebra_appendix} for more detail.}. 
I.e., it is an $SU(N)$ Kac-Moody algebra at level $2N$ acting on fluctuations of the Fermi surface, in the limit of $K\rightarrow\infty$ (and $m,n$ fixed).

With these definitions, $\dD_2$ can be put into a suggestive form: 
\begin{equation}\label{dilatationop}
\dD_2  =
    g^2 \left\lbrace 2 Q + \frac{1}{N}\sum_{u=1}\frac{1}{u} J^a_{-u} J^a_{u}  \right\rbrace. 
\end{equation}
where the number operator $Q=\sum_{k=0}^\infty \rho^a_{k}\check\rho^a_{k} $ is just the U(1) charge, which we turn on to populate the Fermi surface.  

We  immediately see that this opens up a gap of $\Op(g^2)$ between states which are annihilated by $J^a_u$ with $u>0$ and those which are not. Although  (\ref{dilatationop}) has an explicit factor of $N$, it goes away when acting on gauge invariant states.  
%This can be seen explicitly in the computation for generic single-trace fluctuations in Appendix \ref{appendix_d2_fluct}. 
We will come back to this in next subsection. 

%%%%%%%%%%%%%%%%%%%%%%%%%%%%%%%%%%%%%%%%%%%%%%%%%%%%%%%%%%%%%%%

\subsection{Emergence of $SU(N)_{2N}$ gauging}
Before, we focused on ${\cal H}_F$ which are small fluctuations around the fermi surface of the form $SU(N)_N\times SU(N)_N \over \text{global } SU(N)$.  More precisely, we will take ${\cal H}_F$ to be excitation of the Fermi surface  within a band $(K- s , K + s)$ in a fixed $s$ and large $K$ limit. States in ${\cal H}_F$ will be denoted by $\ket F$.

Guided by the form of $\dD_2$, we will further divide ${\cal H}_F$ into
\begin{itemize}
\item ${\cal H}_L$ which includes light states, which satisfy $J^a_u \ket{L} = 0$, for $u > 0$. These are all the primaries of the Kac-Moody algebra.
\item ${\cal  H}_H$ which is its orthogonal complement. It contains heavy states, for which $J^a_u \ket{H} \ne 0$, for $u > 0$. More precisely, given the hermiticity properties of $J$, ${\cal H}_H$ are all descendants (of states in ${\cal H}_L$) under the $SU(N)_{2N}$ Kac-Moody symmetry.
\end{itemize}

It is clear that the states in ${\cal H}_L$ are nothing but the states of the "strange metal" gauged model
\begin{equation}
SU(N)\times SU(N)\over SU(N)_{2N}
\end{equation}
For these states the anomalous dimensions $\delta \D_2$ is just proportional to the charge, which we can shift away by renormalizing the $U(1)$ charge, with the net result that the energy of these states is the same as the classical energy. The remaining states, i.e. those in ${\cal H}_H$ receive another correction which is proportional to $\Op(g^2)$. This separation of scales allows us to truncate our theory to ${\cal H}_L$ alone. 

The cancellation of $\delta \D_2$ is unusual, but a non-vanishing, order $g^2\times \Op(1)$ correction is typical of generic operators. We therefore expect that as we increase $g^2$, to go to the strong coupling limit, states in ${\cal H}_H$ will receive a large anomalous dimension. The rest of the paper is devoted to the issue of whether ${\cal H}_L$ remains massless, in the large $K$ limit. We will see that this true also to order $g^4$.

%%%%%%%%%%%%%%%%%%%%%%%%%%%%%%%%%%%%%%%%%%%%%%%%%%%%%%%%%%%%%%%
\section{Diagrammatics at large $K$}
%%%%%%%%%%%%%%%%%%%%%%%%%%%%%%%%%%%%%%%%%%%%%%%%%%%%%%%%%%%%%%%

Our goal is to push the calculations beyond $\Op(g^2)$, which is challenging since the number of loops increases rapidly, and spin chain techniques are not implementable on the fermi surface, at least not naively. We do expect simplification at large $K$, so we would like to  systematically develop the diagrammatics in this limit. First we take the large $K$ continuum limit, and then deduce the rules for diagrammatics. To check our diagrammatics, we
reproduce the $\dD_2$ result (at the end of this section). Using these techniques we then compute the $\dD_4$ (in the next section).

%%%%%%%%%%%%%%%%%%%%%%%%%%%%%%%%%%%%%%%%%%%%%%%%%%%%%%%%%%%%%%%

\subsection{Continuum Limit}

%%%%%%%%%%%%%%%%%%%%%%%%%%%%%%%%%%%%%%%%%%%%%%%%%%%%%%%%%%%%%%%

To go to the large $K$ limit, we scale the quantities above as follows
\begin{eqnarray}
\rho^a_q   \rightarrow { \rho^a(x) \over \sqrt{K} },\ \ \ \check \rho^a_q &\rightarrow &  { \check  \rho^a(x) \over \sqrt{K} },\ \ \{ \rho^a(x) , \check \rho^b(y) \}  = \delta(x-y) \delta^{ab} \\
q = xK,\ \ \ \sum_k &\rightarrow & K \int_0^\infty dx
\end{eqnarray}
$q$ takes non-negative integer values, where as $x$ is a non-negative real number in the large $K$ limit. 

Before, in $\dD_2$ we had an expression $\sum_{u=1}^{\infty} \frac{1}{u} J^a_{-u} J^a_{u}$ which in the continuum, as we will see, goes over to 
\begin{equation}
\dD_2\sim \int \frac{1}{z} J^a(-z) J^a(z) dz.
\end{equation} 
We need to be careful about the lower limit of integration, which started its life as the $u=1$ term in $\dD_2$. The latter maps to $z=u/K\rightarrow 0$, leading to an apparent singularity of the integrand at $z=0$. In fact, much of our discussion is anchored at such singularities. Similarly, at some place we will need to distinguish momenta factors like $q+1$ from $q$. To do so, we introduce 
\begin{equation}
\epsilon={1\over K}
\end{equation}
We will treat $\epsilon$ as a cut-off of low momenta of the fermion, and introduce it only when divergences appear. Note that although from the basic fermion point of view this is an IR quantity, it will actually be, for the most part, a UV quantity from the point of view of fluctuations around the fermi surface. The reason for this is that it is associated with momenta of fermions and holes very far from the edge of the fermi surface and hence these are high energy states as far as states in $|F\rangle$ are concerned. 

With these rescaling, the previously defined current $J$ has a finite  limit
\begin{eqnarray}
J^a(x_1) &=& -i f^{abc} \int dx_2 \sqrt{   x_2 \over x_1 + x_2 } \rho^b(x_2) \check \rho^c(x_1+x_2) \\
J^a(-y_1) &=& - i f^{abc} \int dy_2  \sqrt{ y_2 \over y_1 + y_2} \rho^b(y_1+y_2) \check \rho^c(y_2) 
\end{eqnarray} 
%With these definition of currents, one can check that $[J^a(x)]^\dagger = J^a(-x)$.  
The $I$ operators  also have nice continuum limit:
\begin{equation}
\label{Iplus}
I_1^+ = {i \over 2 \sqrt{2} } f^{abc}  \int dx_1 dx_2\ \sqrt{x_1+ x_2 \over x_1 x_2 }\rho^a(x_1) \rho^b(x_2) \check \rho^c(x_1+x_2 )  
\end{equation}
\begin{equation}
\label{IMinus} 
\bar I_1^- = {i \over 2 \sqrt{2} N } f^{def}  \int dy_1 dy_2 \ \sqrt{ y_1 + y_2  \over y_1 y_2  }\rho^d(y_1 +y_2) \check \rho^e(y_2 ) \check \rho^f(y_1) 
\end{equation} 
and finally, in these terms, the one-loop dilatation operator $\dD_2$ is given by
\begin{equation}
\dD_2 = g^2 \left( 2 \int_0^\infty dx \rho^a(x) \check \rho^a(x) + 
\frac{1}{N} \int_{\epsilon}^1 \frac{dz}{z} J^a(-z) J^a(z) + {\mathcal O} (1/K) \right)
\label{d2cont}
\end{equation}

%%%%%%%%%%%%%%%%%%%%%%%%%%%%%%%%%%%%%%%%%%%%%%%%%%%%%%%%%%%%%%%%%%%%

\subsection{Singular and Regular Operators}\label{Argument}

%%%%%%%%%%%%%%%%%%%%%%%%%%%%%%%%%%%%%%%%%%%%%%%%%%%%%%%%%%%%%%%%%%%%

Expression \eqref{d2cont} is made out of two distinct terms. Both are integrals (over momenta) of some momenta dependent operators. In the 2nd term, however, there is an additional dependence on the momenta $z$ and, furthermore, this dependence is naively singular at $z=0$ as it goes like $1/z$. We will refer to the first term as {\it regular} and to the last term as {\it singular}. More generally, as we go to higher loops the Hamiltonian can be written as a sum over more and more complicated terms of the form 
\begin{equation}\label{anyop}
\Op_{m,f}=  \int dx_1 ... dx_{2m} \ \delta(x_1+...-x_{m+1}...) \ f_m(x_1,...,x_{2m}) \ :\rho(x_1)..\rho(x_m) \check \rho(x_{m+1})..\check \rho(x_{2m}):. 
\end{equation}
and we can divide the terms into singular or regular depending on whether $f$ has a singularity at some value of the $x's$. 

Rephrasing the discussion above for $\dD_2$, we consider a state in $\ket{F} \subset {\cal H}_F$, with particles and holes in an interval $\delta \ll 1$ around the fermi surface in the large $K$ convention\footnote{I.e., momenta $s \equiv \delta K$ from the fermi surface in the discrete convention.}. The energy of a generic fluctuation state $\ket{F}$, as shown by an explicit computation in Appendix \ref{appendix_d2_fluct}, is $\sim g_{ym}^2 \log \left(\delta\over\epsilon\right)$. This is what we expect from the $1/z$ pole, and it is a contribution which remains finite in the $K\rightarrow\infty$. Note that the powers of $N$ cancel. On the other hand, states in $\ket{L}$ remain at zero energy, up to powers of $1/K$. Note that if we had a contribution of the form, say, $\int JJ$ with no $1/z$ pole in the integrand, then the result would be of $\Op(\delta)$, and would be zero at the large $K$ limit.

Given this terminology we see that the singular operator in $\dD_2$ is responsible for creating the gap between the states of $SU(N)_N \otimes SU(N)_N$ and the gauged model $SU(N)_N \otimes SU(N)_N/SU(N)_{2N}$. Of course, the term is not really singular, since we cut of the integral at $z>\epsilon$. The regular term can't close the gap. In this subsection we argue that this is the general case - only singular terms can create or close the gap and regular terms can only bring about small shifts in each band. This simplifies the perturbative computation considerably since if we want to establish the existence of a gap at higher order in perturbation theory we need to track only the singular composite operators. 

In the following sections we will track the singular pieces in $\dD_4$ in the limit $K\rightarrow\infty$ and show that they vanish on states in ${\cal H}_L$. Hence the gap is not closed also at order $g^4$, and the low energy spectrum is that of the "strange metal".

To show that only singular terms might close the gap, we go back to the expression for ${\cal O}_m$ above and consider the different cases in which the $f_m$'s have or don't have singularities. The fermionic creation and annihilation operators are only those of excitations close to the Fermi surface, i.e, all fermion momenta are within the band $(1 - \delta , 1 + \delta)$. We now proceed to determine the $K$  scaling for each $\Op_{(m)f}$, under the assumption that \eqref{anyop} does not contain any explicit $K$ dependence, in the large K limit, and that such a dependence may show up only via regulating the singularities in the integrand, or when evaluating on states which contain $K$ in them. The reason that this is true to order $\dD_4$ is that $\dD_4$, just as $\dD_2$, can be obtained from commutators and anti-commutator of expressions (such as I's in equation \eqref{Iplus}-\eqref{IMinus}) which are finite in the $K\rightarrow\infty$ limit + $1/K$ corrections. 

To analyze $\Op_{(m)f}$, we will use $\tilde x$ variables defined as $x = 1 + \delta \ \tilde x $. To maintain the canonical commutation relations, this is accompanied by a rescaling $\delta(x) = {\delta(\tilde x) \over \delta}$ and $\rho(x) = {\rho(\tilde x) \over \sqrt{\delta}}$. The operator $\Op_{(m)f}$ can be rewritten as 
\begin{eqnarray}
\nonumber
\Op_{(m)f}&=   \delta^{m-1} & \int_{-1}^1 d\tilde x_1 ... d\tilde x_{2m} \ \delta(\tilde x_1+ ..-\tilde x_{m+1}.. ) \ f_m(1+ \delta \tilde x_1,..,1 +\delta \tilde x_{2m}) \\
&&  \rho(\tilde x_1)..\rho(\tilde x_m)\ \check \rho(\tilde x_{n+1})..\check \rho(\tilde x_{2m}) 
\end{eqnarray}
 Let us also assume that the function $f_m$ has a Taylor expansion :
\begin{equation}\label{fscaling}
f_m(x_1,..x_{2m}) = \sum_{ij}{c_{ij} \over (x_i - x_j)^p}  + \text{less singular}
\end{equation}
for some $i,j$ and $p$ with some constants $c_{ij}$.  Recall that small momenta divergences will be cut off by the regulator $\epsilon$.  Therefore the net scaling of $\bra F  \Op_{(m)f} \ket F$  is  $  \delta^{m-1} \over \epsilon^p$. Hence 
\begin{equation}
\bra F  \Op_{(m)f} \ket F \sim \left({1 \over K}\right)^{m-1-p}
\end{equation}
Hence all operators $\Op_{(m)f}$, with $m- 1 >p $, are $\Op(1/K)$ suppressed and hence vanish in the large $K$ limit. These terms can not close the $\Op(K^0)$ gap between the light states and arbitrary fluctuation states, and only terms with $m-1=p$\footnote{These terms give rise to $K$ independent or $log(K)$ contributions. We will handle them as and when they appear.} or $p>m-1$ are dangerous and should be tracked. This means for a given operator, characterized by $m$, only singular enough integrands can close the gap. For the case of $\dD_2$ above the singular term has $m=2,\ p=1$ and hence it gives rise to a finite gap. 

Note that the above argument shows that a regular two fermion operator do give  a $\Op(1)$ difference between different $\ket F$ states. To maintain the gap, these terms have to be explicitly subtracted away by a chemical potential for $U(1)$ charge operator $Q$. For other $m \ge 2$, regular operators (with $p=0$), can never close the gap. Henceforth, we will use the term \textit{Regular} operator for all operators $\Op_{(m)f}$ with $m -1 > p$.
%%%%%%%%%%%%%%%%%%%%%%%%%%%%%%%%%%%%%%%%%%%%%%%%%%%%%%%%%%%%%%%

\subsection{ A diagrammatic representation for $\dD_2$}\label{d2diagrams}
\begin{figure}[h!]
\centering
\begin{subfigure}[b]{0.45\linewidth}
\includegraphics{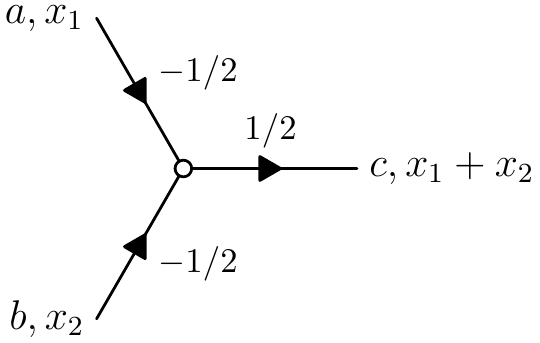}
\end{subfigure}
\hspace{10mm}
\begin{subfigure}[b]{0.45\linewidth}
\includegraphics{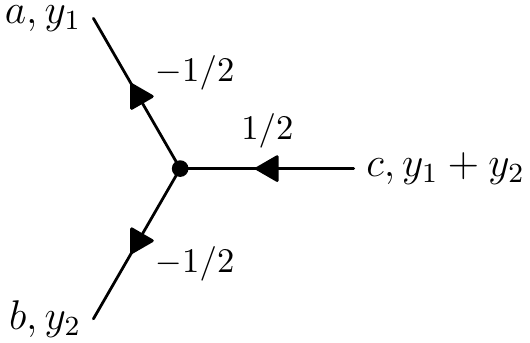}
\end{subfigure}
\caption{Diagrammatic representation of $I^+$ on left and $\bar I^-$ on right, both proportional to $f^{abc}$.}
\label{Iplusminus}
\end{figure}

We now describe a diagrammatic representation for the continuum expressions given in the last section. Using this diagrammatic expansion it turns out that obtaining the singularity structure of $\dD$ is much simpler than the full explicit calculations.  The diagrammatic representation of $I^+$ and $\bar I^-$ is given in Figure \ref{Iplusminus}. 

The expressions for $I^+, \bar I^-$ are given \eqref{Iplus} and \eqref{IMinus}. The incoming (outgoing) arrow indicates a $\rho$ ($\hat \rho$) fermion. Each line is accompanied by a number which indicates the power of momenta that goes along with this line. Additional vertices will be introduced later when we compute higher orders in $g$.

We would like to compute $\dD_2 \sim \{ I^+, \bar I^- \}$, which means, nominally, the contraction of a single line between these two vertices. The expression that we are after, however, is one in which we have only fermion creation and annihilation operators that have momenta in the shell $(1-\delta, 1+\delta)$. This means that we need to contract additional lines which have momenta outside this shell. We will therefore obtain 4-fermion terms with a single line contraction and two-fermion terms with a 2 lines contracted.  

There are additional rules in how to tie the different lines, associated with the ordering of the fermions and of the vertices, and then with how we apply them to the states:
\begin{itemize}
\item $I^+$ is to the left of $\bar I^-$. This is true just because of the fact that $I^+ \ket F  =0$. 
\item The left outgoing arrow on the vertex on right (i.e $\bar I^-$), always has momenta $\le 1+\delta$. This is true because $\check \rho(x)$ to the right will annihilate the $\ket F$ unless $x <1$. A similar reasoning shows that right ingoing arrow on a vertex on the left will have momenta $\le 1$. 
\end{itemize}
Using these rules it is a straightforward, if somewhat laborious, to enumerate all the possible diagrams. The situation simplifies somewhat when we take into account the fact that we are interested only in singular terms. 

\begin{figure}[h!]
\centering
\begin{subfigure}{.5\textwidth}
\centering
\includegraphics{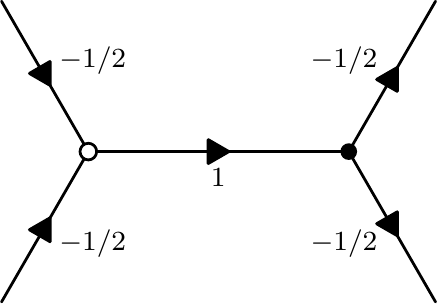}
\end{subfigure}%
\begin{subfigure}{.5\textwidth}
  \centering
  \includegraphics{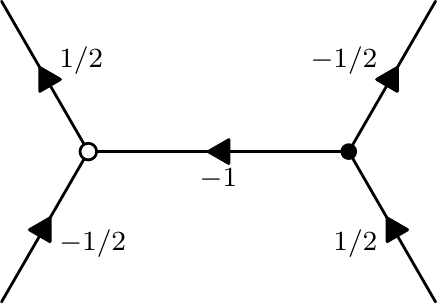}
\end{subfigure}
\caption{Diagrams representing four-fermion operators in $\dD_2$.}
\label{D2tree}
\end{figure}

\medskip
{\bf Four fermi terms}: First consider those terms in which all four of the fermions have momenta near the fermi surface. Diagrams for such terms will have four external legs which are given in Figure \ref{D2tree}.
The diagram on the right has an internal line with weight (-1), in which momenta close to zero can flow. 
Hence it will give a singular contribution. However, it is clear that in this limit the left hand and right 
hand vertices will each give a current algebra generator at the same low momenta. 
Explicitly, the diagram evaluates to
\begin{eqnarray}
\nonumber
&&f^{abe} f^{cde}\int dx_1 dx_2 dx_3 dx_4 \delta(x_1 + x_2 - x_3 - x_4) \rho^b(x_2) \check \rho^a(x_1) \rho^d(x_2) \check  \rho^c(x_3 )  {1 \over x_1 - x_2} \sqrt{ x_3 x_1\over x_4 x_2 } \\
&\sim&   \int  {du \over u} J^a(-u) J^a(u)
\end{eqnarray}
which is the singular term in $\dD_2$ which we identified before.
The diagram on the left is not singular and therefore does not interest us in the large $K$ limit.

\medskip

\begin{figure}[h!]
\centering
\begin{subfigure}{.5\textwidth}
\centering
\includegraphics{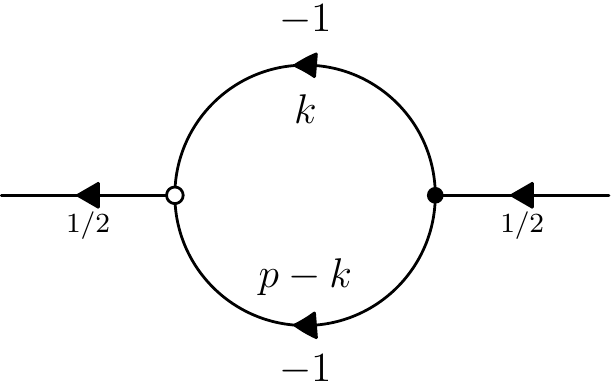}
\end{subfigure}%
\begin{subfigure}{.5\textwidth}
  \centering
  \includegraphics{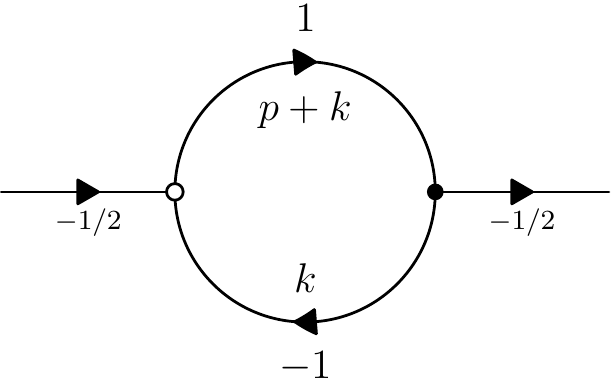}
\end{subfigure}
\caption{Diagrams representing two-fermion operators in $\dD_2$.}
\label{D2loop}
\end{figure}

{\bf Two fermi terms}: To get a two fermion operator in $\dD_2$, we need one more contraction. There are two possibilities as given in Figure \ref{D2loop}.
The left diagram evaluates to 
\begin{eqnarray}
\label{1loopcalc}
N \int dp &  \rho^a(p)  \check \rho^a(p) \ p \ \int_{\epsilon}^{p -\epsilon}  dk {1 \over k (p - k)} = 2 N  \int dp   \rho^a(p)  \check \rho^a(p) \log\left({p \over \epsilon} -1 \right) \nonumber \\
& =  2 N  \log(K) \int dp   \rho^a(p)  \check \rho^a(p)   + \Op(1 / K)
\end{eqnarray}
and the right diagram evaluates to
\begin{eqnarray} 
& N \int \frac{dp}{p}   \rho^a(p)  \check \rho^a(p) \int_{\epsilon}^1 dk {p+ k \over k } = N \int dp   \rho^a(p) \check \rho^a(p) \left[ \log(K) + 1 \right]   + \Op(1 / K)
\end{eqnarray}
where the external momenta are of order $1$ and $\epsilon=1/K$.

These diagrams can be neglected, actually, for multiple reasons, some of which will generalize to higher loops as we will later on.
\begin{itemize}
\item When computing their coefficients more carefully, one sees that the term proportional to $log(K)$ cancels between the two diagrams.
\item Terms which are of the form $F(K)\int d\rho \rho^a(p)\check \rho^a(p) = F(K)  Q$ can be absorbed into a renormalization of the chemical potential.
\item If the integrand which multiplies $\rho^a(p){\check\rho}^a(p)$ has initially a momentum dependence, as is the case here, then when expanding this momenta around $1$, it leads to a term which is proportional to the charge, as in the item before, up to $O(1/K)$ terms which we neglect.
\end{itemize}

For the case of $\dD_2$, this is a verification that the large K diagrammatic technique is useful for rapidly extracting the singular pieces, which are our main interest. In the next section we will apply the same diagrammatic rules to obtain the singular pieces in $\dD_4$.

%%%%%%%%%%%%%%%%%%%%%%%%%%%%%%%%%%%%%%%%%%%%%%%%%%%%%%%%%%%%%%%
\section{Higher orders in Perturbation Theory}
%%%%%%%%%%%%%%%%%%%%%%%%%%%%%%%%%%%%%%%%%%%%%%%%%%%%%%%%%%%%%%%
In this section we consider $\dD_4$, the $\Op(g^4)$ correction to the 
dilatation operator. Again, we find that the gap between the light states and 
generic fluctuations persists. We now briefly summarize the results of this study. 

This operator contains six-fermion, four-fermion and two fermion-terms.  
The six-fermion diagrams, as we show in \ref{six_fermion_sect}, are non-singular in the 
sense of the previous section, and thus do not close the gap. 
From the four-fermion terms, 
the only type of singularity we encounter is (to be made more precise below) of the form 
\begin{equation}
\Op_{4f} \approx \log K \int du \frac{1}{u} {J}^a(-u) {J}^a(u)   
\end{equation}
so that these terms are consistent with the gap found in $\dD_2$. The two fermion terms does not contain any singular pieces, and, up to $1/K$ corrections, is proportional to charge, and can be shifted away by a redefinition of the chemical potential (see Appendix \ref{appendix_D4_2fermion} for details).

In section \ref{D_4_computation_sect} we present the all order ansatz for the $\psu(1,1|2)$ sector and its application to the fermionic $\su(1,1)$ sector. Application of this procedure allows us to compute $\dD_4$ in the large $K$ limit. In the process we will need some additional vertices, on top of the ones that we already discussed, and in section \ref{D_4_cont_limit} we describe their continuum limit and the resulting Feynman rules. Finally, the diagrams are evaluated in sections \ref{four_fermion_sect} and  \ref{six_fermion_sect}. 

\subsection{Computation of $\delta {\cal D}_4$}
\label{D_4_computation_sect}
In this section we compute $\delta {\cal D}_4$, and we would like, eventually, to have an all order proof. We will therefore describe $\delta {\cal D}_4$, after a short digression for the suggested all order ansatz for this sector described in \cite{Zwiebel:2008gr}.

First, observe that the Next-to-Leading-Order correction to $I^+$, $\bar{I}^-$, and in fact every NLO correction to the $\mathfrak{psu}(1,1|2)$ generators, is given by the following schematic form
\begin{align}
J_{NLO} &= \pm \lambda [ J_{LO}, \mathfrak{X} ]\\
\mathfrak{X}&=\frac{1}{2} \epsilon_{ab} \left\{ Q^b_{LO}, \left[ S^a_{LO},h \right] \right\} + h.c.
\end{align} 
where $\lambda$ is the 't Hooft coupling $\lambda = 16 \pi^2 g^2$. 
The sign depends on whether the generator corresponds to a positive or negative Lie algebra root.
Here, $h$ is an axillary generator, which is just the harmonic generator (\ref{harmonic_operator}) at zeroth order in $\lambda$
\begin{align}
   h  =& \sum_{n=0}^\infty
     \frac12h(n+1) \left( \tr\cno{\psi_{(n)}\check \psi_{(n)}}+\tr\cno{\bar\psi_{(n)}\check{\bar\psi}_{(n)}} \right)
 +\nonumber \\ & \sum_{n=0}^\infty \frac12h(n)\sum_{i=2}^3\tr\cno{\phi^i_{(n)}\check \phi^i_{(n)}}~ 
 \label{harmonic_operator}.
\end{align}

In \cite{Zwiebel:2008gr}, Zwiebel conjectured that this type of structure continues to all orders. This can be realized 
by replacing the above equation by
\begin{equation}
\frac{\partial}{\partial\lambda} J(\lambda) = \pm \lambda [ J(\lambda) , \mathfrak{X}(\lambda) ].
\end{equation}
where now 
\begin{equation}
\mathfrak{X}(\lambda) = \epsilon_{ab} \left\{ Q^a(\lambda), \left[ S^b(\lambda), h(\lambda )\right] \right\} + \frac{1}{2} \left[ H(\lambda), h(\lambda) \right]
\end{equation} 
with $\lambda H(\lambda) = \dD(\lambda)= 2 \left\{ I^+(\lambda),\bar{I}^-(\lambda) \right\}$. $h(\lambda)$ generalizes the harmonic generator. One can find this generator by solving using equation (3.3) in \cite{Zwiebel:2008gr}, which must be obeyed in order to preserve the Lie algebra symmetry constraints. 
One can now solve these equations for $J(\lambda)$ and $\mathfrak{X}(\lambda)$ order by order in $\lambda$. 
This proposal has been used to compute $\dD_6$, which passes some non-trivial tests.

%%%%%%%%%%%%%%%%%%%%%%%%%%%%%%%%%%%%%%%%%%%%%%%%%%%%%%%%%%%%%%%%%%%
%\subsubsection{Computation of $\dD_4$}
%%%%%%%%%%%%%%%%%%%%%%%%%%%%%%%%%%%%%%%%%%%%%%%
Although the iterative procedure outlined in the above section provides a way to compute the anomalous dimensions to any order, in this work we restrict to $g^4$ order in perturbation theory. 
Whether by using the iterative procedure above, or by a direct computation in ${\cal N}=4$ SYM as in \cite{Zwiebel}\footnote{Conventions of \cite{Zwiebel} are related to ours by $\overrightarrow{I}^{\pm} \rightarrow I^\pm$ and $\overleftarrow{I}^{\pm} \rightarrow \bar I^\pm$}, the expression for $\dD_4$ is
\begin{equation}\label{strctd4}
\dD_4 = 2 \left\{ \bar I_1^- , \left[I_1^+,\left\{I_1^-, \left[ \bar{I}_1^+, h \right] \right\} \right] \right\} + 
2 \left\{  I_1^+ , \left[\bar I_1^-,\left\{\bar I_1^+, \left[ {I}^-_1,  h \right] \right\}\right] \right\}.
\end{equation}
where the expressions for $I_1^\pm ,{\bar I}_1^\pm$ in the full $\psu(1,1|2)$ sector are given in Appendix \ref{d4details}  (\ref{supercharges}).  It is convenient to define
\begin{eqnarray}
\label{Vdef}
2 \{ \bar I_1^+ , [I_1^- , h] \}  \equiv V\ \ \\
\label{Cdef}
2 \{ I_1^-, [\bar I_1^+ ,h] \}  \equiv C\ .
\end{eqnarray}
With this, $\dD_4$ can be rewritten as 
\begin{equation}\label{expd4}
\dD_4 =  \{ \bar I_1^- , [I_1^+ , C] \}  +  \{ I_1^+ , [\bar I_1^- , V] \}   = I_1^+ (C -V) \bar I_1^- - C I_1^+ \bar I_1^- + I_1^+ \bar I_1^- V
\end{equation}
In going to second line we have dropped $I^+$ ($\bar I^-$) acting on right (left). The expressions for $V,C$ (as computed in the Appendix \ref{d4details}) are 
\begin{eqnarray}
\label{Vexp}
V &=& {i \over 2 N} \sum_{m=0 , u =1}^\infty B_{m,u} \ f^{abc} \rho^b_{m+u} \check \rho^c_m \ J^a_u +  \sum_{m=0}^\infty B_m  \rho^a_m \check \rho^a_m = V_{4f}+V_{2f}\\
\label{Cexp}
- C &=&  {i \over 2 N} \sum_{q =0 , u =1}^\infty B_{q,u} \ J^a_{-u} \ f^{abc}  \rho^b_q \check \rho^c_{q+u} +  \sum_{m=0}^\infty B_m  \rho^a_m \check \rho^a_m=-C_{4f}-C_{2f}.
\end{eqnarray}
where 
\begin{eqnarray} 
B_{m,u} &= \sqrt{m+1\over  m+u+1 } \  {h(m+u+1) - h(m+1) - h(u)  \over u } \\ 
B_m  =&  h(m+1) - 2.
\end{eqnarray}
In $\bra L  I_1^+ \bar I_1^-  V \ket L \supset \bra L \dD_4 \ket L $, only the two fermion part of $V$ contributes. This is because the four fermion part of $V$ has $J_u$ on the right which annihilates $\ket L$. Similarly one can see that only the two fermion part of $C$ contributes to $- C I^+_1 \bar I^-_1$. It is convenient to define 
\begin{equation}
U = C-V
\end{equation}
whose four fermion part will be called $U_{4f}$ and the  two fermion part $U_{2f}$. Since $I^+ \bar I^- \ket L \sim Q \ket L$ and $Q$ commutes with the two fermion part of $C,V$, one can further simply the expressions for  $\dD_4$ to get
\begin{equation}\label{dD4simp}
\dD_4 = I^+ (U_{2f} + U_{4f}) \bar I^-  - U_{2f} I^+ \bar I^-  = I^+ U_{4f}  \bar I^- + \{ [ I^+ , U_{2f} ], \bar I^- \}
\end{equation}

%%%%%%%%%%%%%%%%%%%%%%%%%%%%%%%%%%%%%%%%%%%%%%%
\subsection{$\dD_4$ in the continuum limit}\label{D_4_cont_limit}
%%%%%%%%%%%%%%%%%%%%%%%%%%%%%%%%%%%%%%%%%%%%%%%
We can now take the continuum limit of the above expressions. It is useful to define another small parameter $\tilde \epsilon$ which arises in this limit,
\begin{equation}\label{teps}
\tilde \epsilon =  {1 \over \log(K e^\gamma) }
\end{equation}
The continuum limit of the operators $C,V$ are 
\begin{eqnarray}\nonumber
 C &=& { i \over 2 N \tilde \epsilon} \int {  dz_2 dz_3  \over   z_3 }  f^{ecd} \sqrt{  z_2 \over z_2+z_3 }   \left( 1+  \tilde \epsilon \log[ {z_2   z_3 \over z_2 + z_3} ]    \right) J^e(-z_3)     \rho^c(z_2)  \check  \rho^d(z_2+z_3)    \\
&&     \hspace{10mm} - {1 \over \tilde \epsilon} \int dx \rho^a(x) \tilde \rho^a(x) \left[1 +  \tilde \epsilon \log(x/2) \right] \\
\nonumber
  V &=& - {i \over 2N \tilde \epsilon }\int {dz_1  dz_3 \over z_3 } f^{eab} \sqrt{z_1  \over z_1+z_3}   \left( 1+  \tilde \epsilon \log[ {z_1 z_3 \over z_1 + z_3} ]  \right)  \rho^a(z_1+z_3) \check \rho^b(z_1) J^e(z_3)    \\
&&  \hspace{10mm}  +  {1 \over \tilde \epsilon} \int dx \rho^a(x) \tilde \rho^a(x) \left[1 +  \tilde \epsilon \log(x/2) \right] 
\end{eqnarray}
The two and four fermion parts of $U$ become, to first order in $\tilde\epsilon$
\begin{eqnarray}
U_{2f} &=& - {2 \over \tilde \epsilon}  \int dx \rho^a(x) \tilde \rho^a(x) \left(x \over 2 \right)^{\tilde \epsilon}+ \Op(\tilde \epsilon )\\  
\label{U4exp}
U_{4f}  &=&   - {1\over N \tilde \epsilon} \int { dz_3 \over z_3^{1 - \tilde \epsilon}} \mathbb{J}^a(-z_3 )\mathbb{J}^a(z_3 ) + \Op(\tilde \epsilon )
\end{eqnarray}
where we have defined a new current 
\begin{equation}
\mathbb J^e(z) = - i \int dz_2  f^{ecd}  \sqrt{z_2^{1 +\tilde \epsilon}} \rho^c(z_2)  {\check \rho^d(z_2+z)   \over \sqrt{(z_2 + z)^{1 + \tilde\epsilon}}}.
\end{equation}

$J$ and $\mathbb J$ have almost same action on fluctuations
\begin{eqnarray} \nonumber
(J^a(z) - \mathbb{J}^a(z) ) \ket F &=&\left(- {i \tilde \epsilon z \over 2}   \int_{1 - \delta}^{1+ \delta- z} { dz_2  f^{acd} \over z_2}   \rho^c(z_2)  \check \rho^d(z_2+z) + \Op(\tilde \epsilon^2 z^2) \right) \ket F \\\label{JJdiff}
&\le & \Op(\tilde \epsilon \delta) \ket F
\end{eqnarray}
Since the new current has the same action on fluctuations (up to $1/K$ corrections), it provides an equally good definition of light states. To see this more clearly, consider $\dD_2$ (as a matrix between fluctuation states) now in terms of $\mathbb J$ 
\begin{equation}
\dD_2 = 2 \int dz \rho^a(z) \check \rho^a(z) + {1 \over N} \int_\epsilon^{2 \delta }  dz \left[  {\mathbb{J}^a(-z  )\mathbb{J}^a(z) \over z}  + \Op(\tilde \epsilon) \times \text{Regular in z} \right]
\end{equation}
Hence we can also define light states $\ket {\mathbb L}$ as those which have $\mathbb{J}_u \ket {\mathbb L}$   for $u>0$. Henceforth, we will use this definition for light states\footnote{One might worry, now that $\bra H  \dD_2 \ket{ \mathbb L} \ne 0$, whether there could be $\Op(g^4)$ mixing effects when we attempt to diagonalize $\dD_2$, which would then destroy the gap. But one can easily estimate the effect of this mixing to be $\Op(g^4/K^2)$ and hence vanishing at large $K$ }.

\begin{figure}[h!]
\centering
\includegraphics{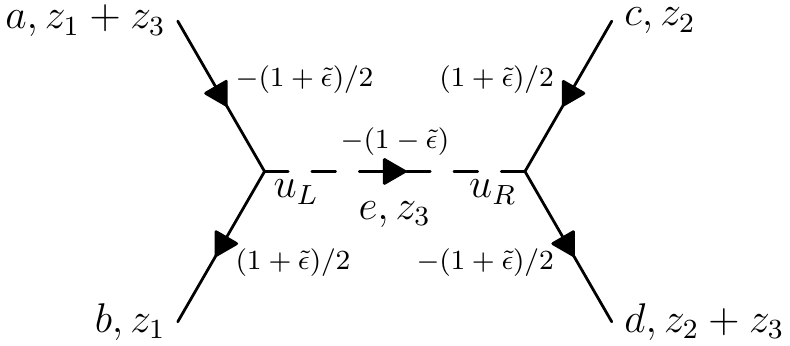}
\caption{Diagrammatic representation of $U_{4f}$, proportional to $f^{abe} f^{cde}$.}
\label{Udiag}
\end{figure}

A diagrammatic representation of the four fermion part of  $U_{4f}$  is given in Figure \ref{Udiag}. We will label the two vertices in the diagram as $U_L,U_R$ as shown in figure.  

%%%%%%%%%%%%%%%%%%%%%%%%%%%%%%%%%%%%%%%%%%%%%%%%%%%%%%%%%%%%%%%

Consider the expression (\ref{dD4simp}) for $\dD_4$, and focus on the term  $\bra  {\mathbb L} I^+ U_{4f} \bar I^- \ket {\mathbb L}$
(the last term $\{ [ I^+ , U_{2f} ], \bar I^- \}$ will be dealt with separately later). The following rules can be applied when evaluating this term 
\begin{itemize}
\item Since $U_{4f} \sim \int {1 \over u} {\mathbb J}(-u) {\mathbb J}(u)$, the expectation value of  $\bra{ \mathbb L} I^+U_{4f} \bar I^- \ket{\mathbb L}$ is just\\
 $\int {du \over u} \bra{ \mathbb L} [I^+, {\mathbb J}(-u)], [{\mathbb J}(u),\bar I^-] \ket{\mathbb L}$.  In terms of diagrammatic representation, 
this means that there is at least one contraction between $I^+, U_L$  and $\bar I^-, U_R$.
\item We can order the diagram so that $I^+, U_L, U_R , \bar I^--$ vertices are in a left to right order. Then all (except the internal line of $U_{4f}$) internal momenta are restricted: All left outgoing arrow (hence right ingoing arrow) on a vertex have momenta $\le 1+\delta$. Also all left ingoing arrow (hence right outgoing arrows) on a vertex have momenta $\ge 1-\delta$. \footnote{ In some diagrams, we will not stick to the convention of ordering the vertices in the diagram from right to left. In this case ordering is assumed to be that $\bar I^-,U_R,U_L, I^+$ acts in a right to left order.}
\end{itemize} 

%%%%%%%%%%%%%%%%%%%%%%%%%%%%%%%%%%%%%%%%%%%%%%%%%%%%%%%%%%%%%%%

\subsection{Six fermion Diagrams in $\dD_4$ between Light States  }
\label{six_fermion_sect}

We classify the diagrams according to the number of loops. It is easiest to start with those diagrams which have no loops. These have all the six fermions close to Fermi surface.  In terms of diagrammatics, they have six external momenta. Also, if any of the diagrams below have a hermitian conjugate counterpart, we don't write it down explicitly since it gives the same contribution. 

\begin{figure}[h!]
\centering
\begin{subfigure}{.5\textwidth}
\centering
\includegraphics[width=.7\textwidth, keepaspectratio]{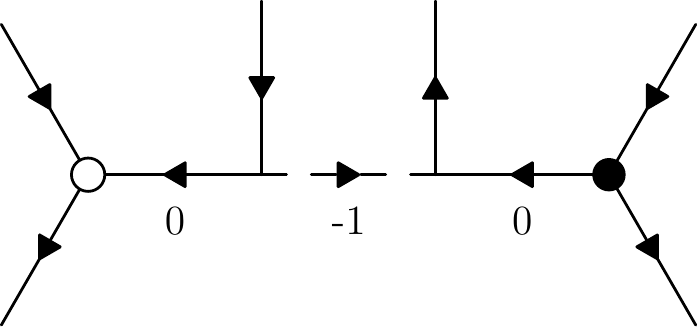}
\end{subfigure}%
\begin{subfigure}{.5\textwidth}
\centering
  \includegraphics[width=.7\textwidth, keepaspectratio]{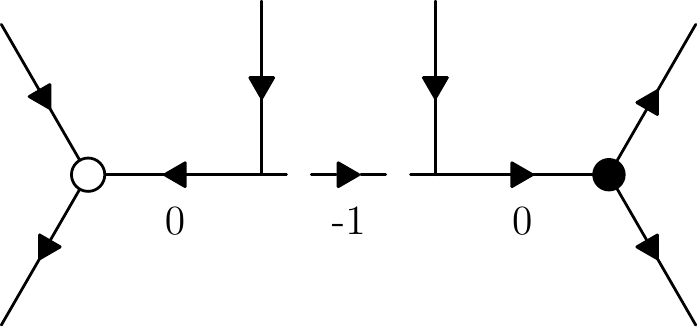}
\end{subfigure}\\
\begin{subfigure}{.5\textwidth}
\centering
  \includegraphics[width=.7\textwidth, keepaspectratio]{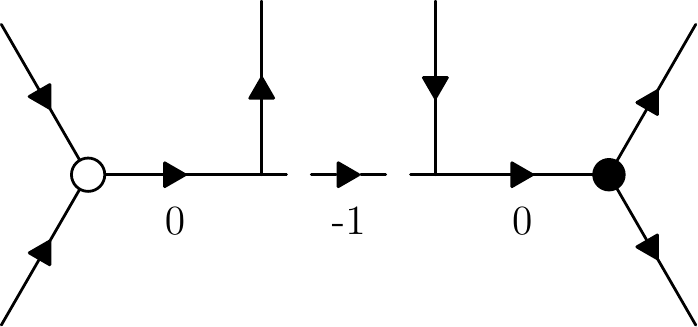}
\end{subfigure}
\caption{Six-fermion operators in $\dD_4$.}
\label{D6tree}
\end{figure}

In Figure \ref{D6tree} we list out all possible diagrams consistent with rules given in the last section. The $\tilde \epsilon$ corrections to momentum degree is not shown in the diagrams because it turns out to be irrelevant for the argument below.

As explained in Section \ref{Argument}, only those terms with a singularity can create a $\Op(1)$ gap. But from the Figure \ref{D6tree}, it is clear that the only line with degree $-1$ has large momenta $\sim 1$ for all the diagrams. Hence these diagrams are  regular and do not create a gap.

%%%%%%%%%%%%%%%%%%%%%%%%%%%%%%%%%%%%%%%%%%%%%%%%%%%%%%%%%%%%%%%

\subsection{Four fermion Diagrams in $\dD_4$ between Light States}
\label{four_fermion_sect}
Next we consider diagrams with one loop. They have four fermions with momenta of $\Op(1)$, i.e four external legs. Before we start computing diagrams, it is useful to investigate the structure of the answer that we expect.

%%%%%%%%%%%%%%%%%%%%%%%%%%%%%%%%%%%%%%%%%%%%

\subsubsection{Comments on general structure of four fermion terms}
It is possible to bring any four fermion term to the following form 
\begin{align}
\Op_{4f} =\int dx_1 dx_2 du & \  \  f^{abg} {\rho^a(x_1+u) \over (x_1 + u)^{1 + \tilde \epsilon \over 2}} x_1^{1 + \tilde \epsilon \over 2} \check \rho^b(x_1) \\ \nonumber
& \ \  f^{cdg} x_2^{1 + \tilde \epsilon \over 2} \rho^c(x_2) { \check \rho^d(x_2 + u) \over (x_2 + u)^{1 + \tilde \epsilon \over 2} } \ F(x_1,x_2,u)
\end{align} 
for some function $F(x_1,x_2,u)$\footnote{Fermions can always be ordered in this way, by using anticommutation relations between them (Any contraction leads to a two fermion term, which we deal separately). Using double line notation, it is also easy to see that the gauge structure can be reduced to the above form.}. Then the singularity structure of this function determines whether the relevant diagram can create a gap. 

\textbf{Singularity structure:} Since the momenta of all the four fermions are $\Op(1)$, we can expand $F(x_1,x_2,u)$ in a Taylor series in $x_1-x_2, u, x_1+x_2-2$. It is convenient to classify terms depending on their scaling with $K$\footnote{ Note that $u, x_1-x_2,x_1+x_2-2$ scale like as $1 \over K$. Functions of form ${x_1 - x_2 \over u}$  scale as $K^0$. }. We find that in all cases, possibly after relabeling the external momenta, the expansion takes the following form \footnote{ The explicit factor of $\log(K \gamma)$ is to keep track of explicit factor of ${1 \over \tilde \epsilon}$ in (\ref{U4exp}).}
\begin{equation}
F(x_1,x_2,u) = \log\left(K e^{\gamma}\right) \lbrace G(u)+ H(x_1,x_2,u) \rbrace \hspace{10mm} 
\end{equation}
where we have classified terms according to their $K$ scaling into a piece $G(u)$ which scales like $K^p$ with $p \sim 1$ (or scales like $\sim {1 \over u}$) and a piece $H(x_1,x_2,u)$ which scales like $K^q$ with $q \ll 1$. As argued in sec  \ref{Argument}, the $H(x_1,x_2,u)$ term does not close the gap and we can drop this from subsequent discussion.

The crucial point now, is that the  part of $F(x_1,x_2,u)$ whose divergence is $\Op(K)$ or worse is independent of $x_1-x_2, x_1 + x_2 - 2$.  This enables us to write
\begin{eqnarray}
\Op_{4f} = \log\left(K e^{\gamma}\right) \int du G(u) {\mathbb J}^a(-u)   {\mathbb J}^a(u)   
\end{eqnarray}  
Since light states satisfy $\mathbb{J}^a(u) \ket {\mathbb L} = \Op({1 \over K}) \ket {\mathbb L}$ for $u>0$ and $G(u)$ only scales like $K$,  $\bra {\mathbb L} \Op_{4f}\ket {\mathbb L}$ vanishes. 
 
For all the diagrams, we now explicitly evaluate the functions $G(u)$ and $H(x_1,x_2,u)$ and find one of the following behaviors: 
\begin{eqnarray}
\label{0reg}
&G( u) = 0&  \hspace{20mm} H(x_1,x_2,u) \sim K^0\\
\label{0log}
&G(u)  = 0&  \hspace{20mm} H(x_1,x_2,u) \sim \log(K) 
\\
\label{rootreghalf}
&G(u) =    {\log(1 + u K)  [1 + \Op(\tilde \epsilon)]  \over  u^{1 - { \tilde \epsilon \over 2}  }   }  &\hspace{20mm}  H(x_1,x_2,u) \sim K^0  \\
\label{rootreg}
&G(u) =    {\log(1 + u K)  [1 + \Op(\tilde \epsilon)]  \over  u^{1 - \tilde \epsilon  }   }  &\hspace{20mm}  H(x_1,x_2,u) \sim K^0 
\end{eqnarray}
Since in all cases, the worst singularity is $G(u) \sim {1 \over u}$, this shows that these diagram cannot close the gap. Note that any power of $\log(K)$ will be considered as weakly $K^0$.

%%%%%%%%%%%%%%%%%%%%%%%%%%%%%%%%%%%%%%%%%%%%
\subsubsection{Explicit evaluation of Four fermion diagrams in $\dD_4$}
We now consider diagrams with four external legs. To get such a diagram, take any of the two figures in Figure \ref{D6tree} and contract any two external legs. It is clear all such resulting diagrams will have one loop and since only planar diagrams contribute  to leading order in $N$, they will be accompanied by a factor of $N$. Note that this cancels the explicit factor of $1/N$ in (\ref{U4exp}). We study the diagrams in increasing order of complexity. 

\begin{itemize}
%%%%%%%%%%%
\item \textbf{External propagator correction:} We begin with the simplest case, which is a correction to the propagator of one of the external legs. Schematically, these are diagrams of the form shown in Figure \ref{external_corr}.
\begin{figure}[h!]
\centering
\includegraphics[scale=0.5]{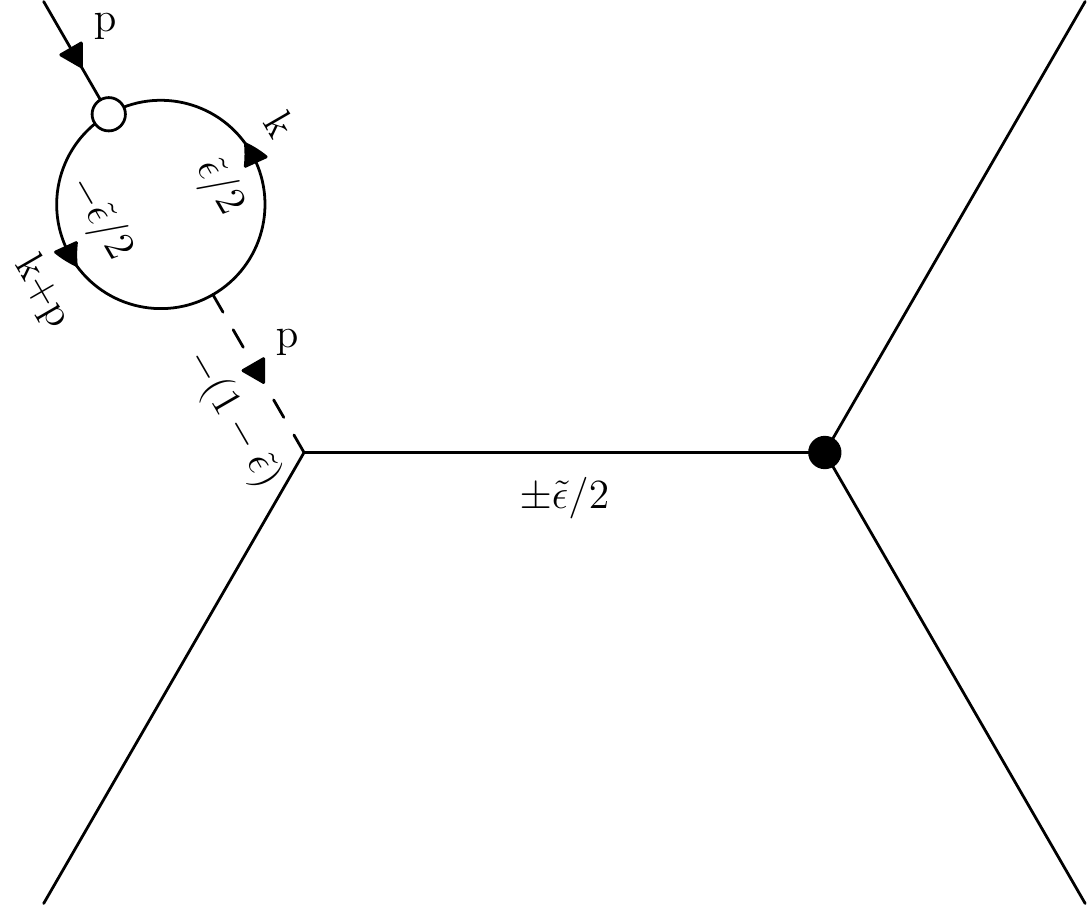}
\caption{Schematic representation of corrections to external legs.}
\label{external_corr}
\end{figure}
Using the rules of previous section, this results in a loop integral 
\begin{equation}
\int_\epsilon^1 dk { k^{\tilde \epsilon \over 2} \over (k+ p)^{\tilde \epsilon \over 2} }  = 1 + \Op(\tilde \epsilon)
\end{equation}
which is of the form given in (\ref{0reg}).

%%%%%%%%%%%
\item \textbf{Internal propagator correction: }Next we consider diagrams with internal leg corrections.  As per rules of  section \ref{D_4_cont_limit}, there are only two diagrams, given in Figure \ref{D4_internal_corr}.
The loop inside the left diagram with $u \sim \epsilon$ evaluates to\footnote{We first extract the $\epsilon$ scaling(in this case $u$ scaling), and then truncate the integral to $\Op(\tilde \epsilon)$.}
\begin{equation}
u^{\tilde \epsilon } \int_\epsilon^{1-u} {dk \over k^{1 - \tilde \epsilon} (k+u)^{1+ \tilde \epsilon}}  = {1 \over u^{1 - \tilde \epsilon}} \log\left(1 +   {u \over  \epsilon}\right) [1 + \Op(\tilde \epsilon) ] +  finite
\end{equation}
This results in functions of the form given by (\ref{rootreg}). 
The loop in the  diagram on the right evaluates to  (for $u \approx 2$). 
\begin{equation} 
u^{-\tilde \epsilon } \int_\epsilon^{u-1} {dk (u-k)^{1+ \tilde \epsilon} \over k^{1- \tilde \epsilon}} =  \log(K) ( finite )
\end{equation}
which is of the form (\ref{0log})
\begin{figure}[h!]
\centering
\begin{subfigure}{.4\textwidth}
\centering
\includegraphics[width=.6\textwidth, keepaspectratio]{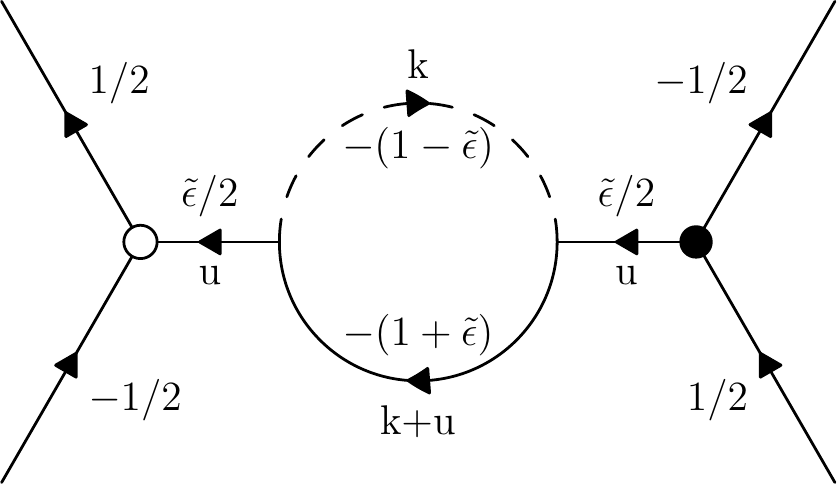}
\end{subfigure}%
\begin{subfigure}{.4\textwidth}
\centering
  \includegraphics[width=.6\textwidth, keepaspectratio]{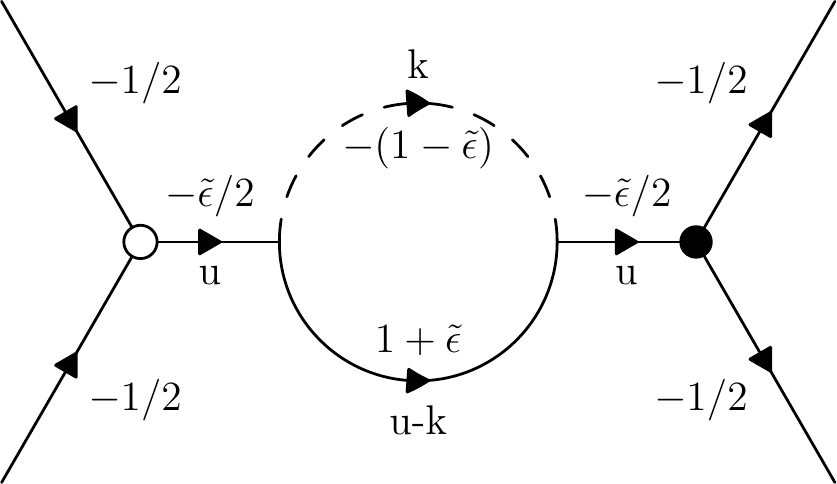}
\end{subfigure}
\caption{Diagrams contributing to internal propagator correction.}
\label{D4_internal_corr}
\end{figure}
%%%%%%%%%%%
\item \textbf{1PI: } Let us now look at the 1PI four-fermion diagrams. 
Using the rules of section \ref{D_4_cont_limit}, there are four possible diagrams given in Figure \ref{D41PIexp}.
\begin{figure}[h!]
\centering
\begin{subfigure}{.4\textwidth}
\centering
\includegraphics[width=.6\textwidth, keepaspectratio]{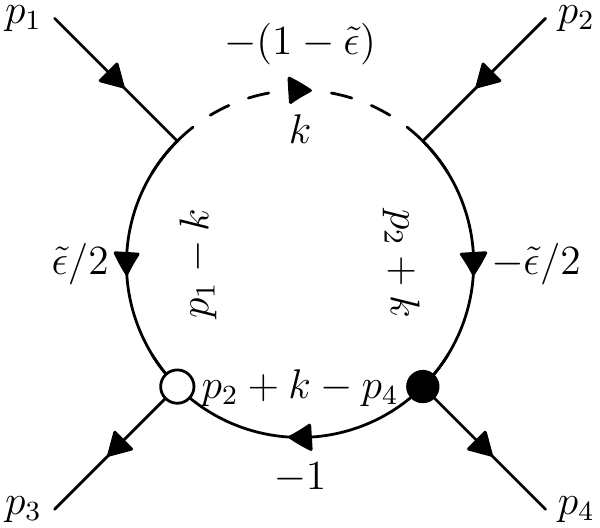}
  \caption{}
\end{subfigure} 
\begin{subfigure}{.4\textwidth}
\centering
  \includegraphics[width=.6\textwidth, keepaspectratio]{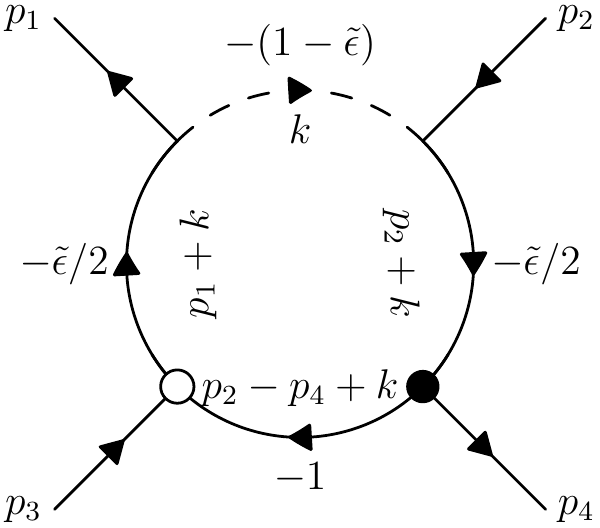}
    \caption{}
\end{subfigure} \\ 
\begin{subfigure}{.4\textwidth}
\centering
  \includegraphics[width=.6\textwidth, keepaspectratio]{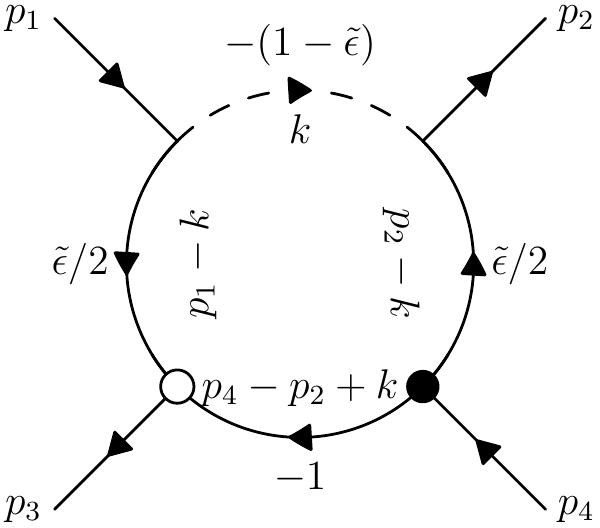}
  \caption{}
\end{subfigure} 
\begin{subfigure}{.4\textwidth}
\centering
  \includegraphics[width=.6\textwidth, keepaspectratio]{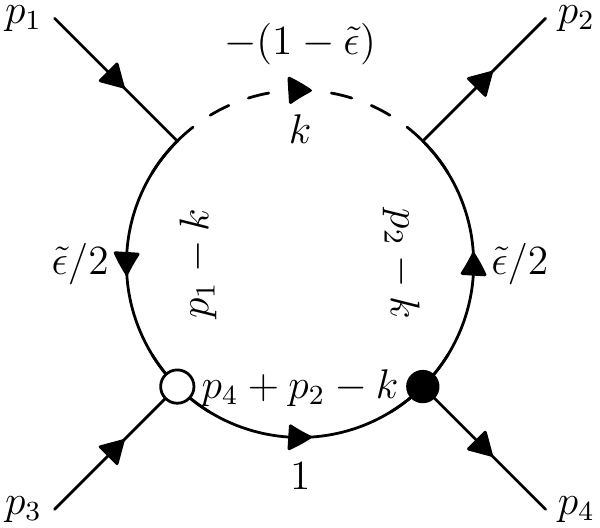}
  \caption{}
\end{subfigure}\\ 
\caption{Vertex corrections}
\label{D41PIexp}
\end{figure}
The result of the loop integral for (a),(b),(c) respectively are
\begin{align}
\int_\epsilon^{1}  & {dk \ (p_1- k)^{\tilde \epsilon \over 2 }    \over (p_2 - p_4  + k) (p_2 + k)^{\tilde \epsilon \over 2 } k^{1 - \tilde \epsilon} } = 
\nonumber \\    & \hspace{5mm}  { \log\left[1+ {p_2-p_4 \over \epsilon}\right] [1 + \Op(\tilde \epsilon)] \over (p_2- p_4)^{1 - \tilde \epsilon }} + \text{finite} 
\hspace{20mm} \begin{array}{ccc} u &\equiv& p_2 - p_4 \\ x_1 &\equiv& p_3 \\ x_2 &\equiv& p_1 \end{array} \\ 
\int_\epsilon^1 & { dk \over   (p_2 - p_4  + k)   (p_1+ k)^{\tilde \epsilon \over 2 } (p_2 + k)^{\tilde \epsilon \over 2 } k^{1 - \tilde \epsilon} } = 
\nonumber \\  & \hspace{5mm} {\log\left[1+ {p_2-p_4\over \epsilon}\right] [1 + \Op(\tilde \epsilon)] \over (p_2- p_4)^{1 - \tilde \epsilon }} + \text{finite} 
\hspace{20mm} \begin{array}{ccc} u &\equiv& p_2 - p_4 \\ x_1 &\equiv& p_1 \\ x_2 &\equiv& p_3 \end{array} \\ 
\int_\epsilon^1 & { dk\  (p_1- k)^{\tilde \epsilon \over 2 }  (p_2 - k)^{\tilde \epsilon \over 2 }\over  (p_4 - p_2  + k)  k^{1- \tilde \epsilon} } = 
\nonumber \\ & \hspace{5mm}  {\log\left[1+ {p_4-p_2 \over \epsilon}\right] [1 + \Op(\tilde \epsilon)]  \over (p_4- p_2)^{1 - \tilde \epsilon }} + \text{finite} 
\hspace{20mm} \begin{array}{ccc} u &\equiv& p_4 - p_2 \\ x_1 &\equiv& p_3 \\ x_2 &\equiv& p_1 \end{array}
\end{align}
All of the above expressions are of the form shown in (\ref{rootreg}).

The result for (d) is
\begin{equation}
\int_\epsilon^1 dk { (p_2- k)^{\tilde \epsilon \over 2 }  (p_1- k)^{\tilde \epsilon \over 2 } (p_4 + p_2 - k) \over k^{1 - \tilde \epsilon} } = \log(K) \times ( \text{finite})
\end{equation}
Correspondingly we get (\ref{0log}).
%%%%%%%%%%%
 \item \textbf{Vertex corrections: }Finally, we analyze the one-loop vertex correction. The relevant diagrams are given in Figure (\ref{D4_vertex_corr}). 
\begin{figure}[h!]
\centering
\begin{subfigure}{.4\textwidth}
\centering
\includegraphics[width=.6\textwidth, keepaspectratio]{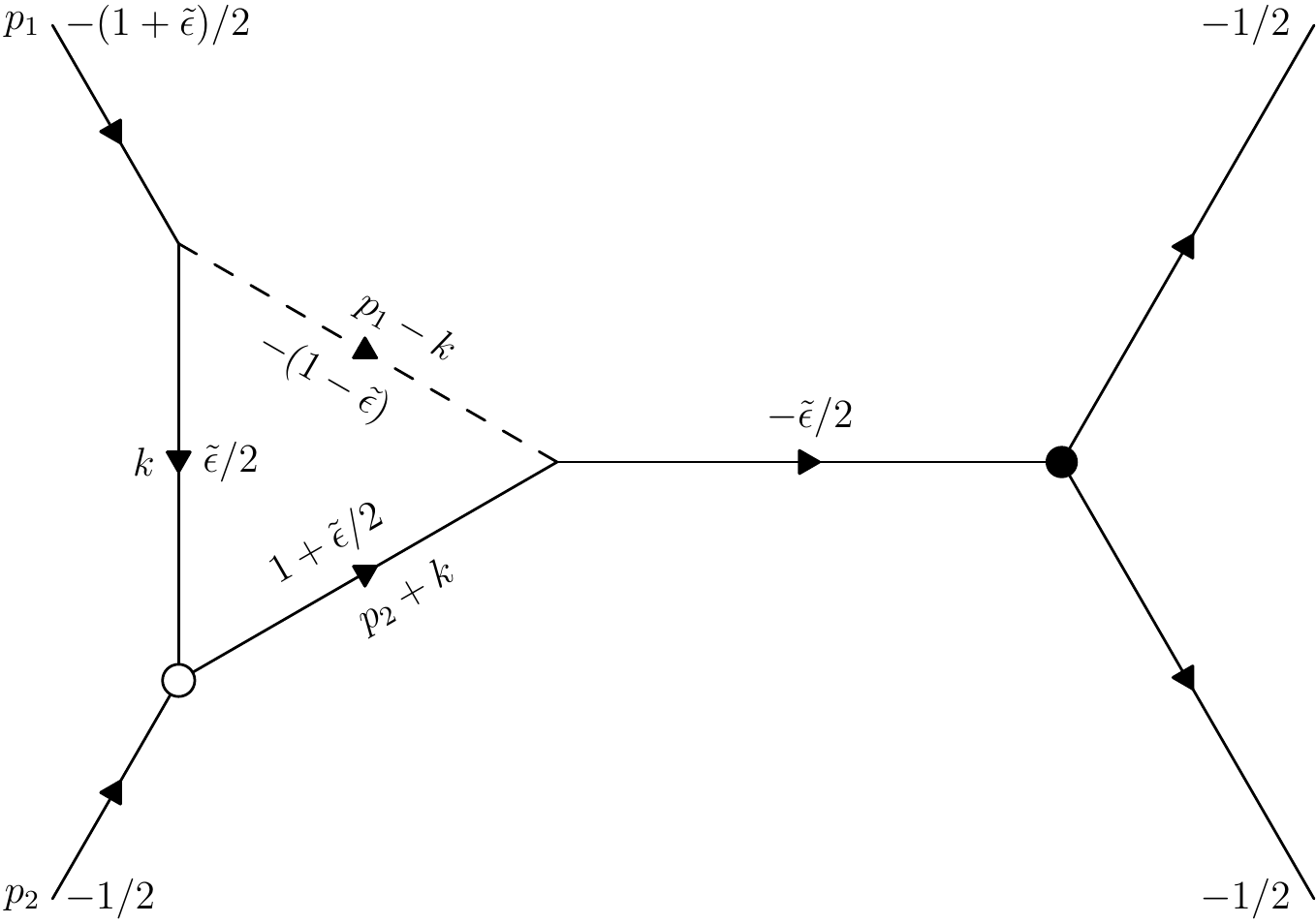}
  \caption{}
\end{subfigure} 
\begin{subfigure}{.4\textwidth}
\centering
  \includegraphics[width=.6\textwidth, keepaspectratio]{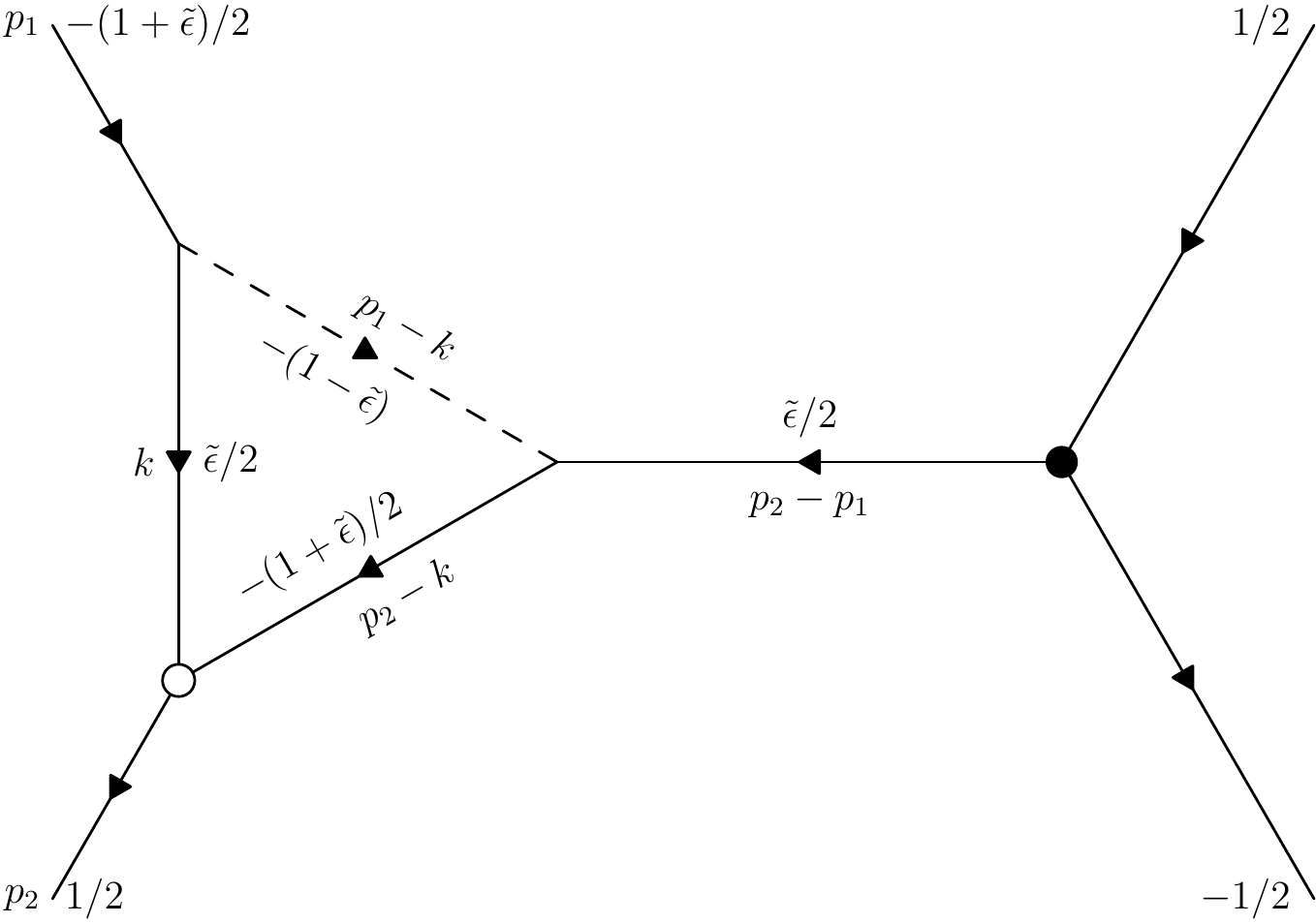}
    \caption{}
\end{subfigure} \\ 
\begin{subfigure}{.4\textwidth}
\centering
  \includegraphics[width=.6\textwidth, keepaspectratio]{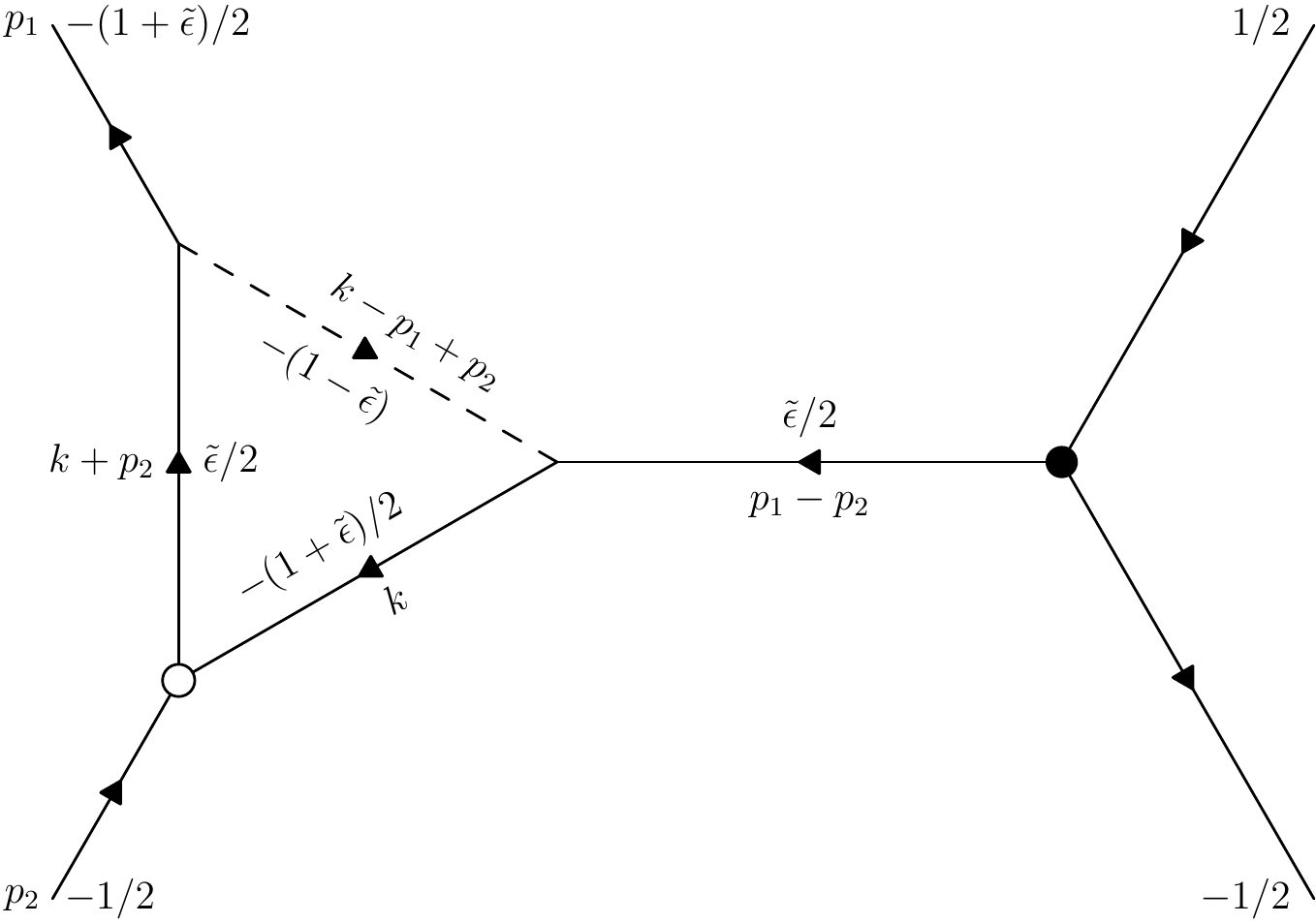}
  \caption{}
\end{subfigure} \\ 
\caption{Vertex corrections}
\label{D4_vertex_corr}
\end{figure}
The loop integrals for (a) is
\begin{equation}
\int_0^{p_1-\epsilon}  dk \ {k^{\tilde \epsilon \over 2} (p_2 + k)^{1 + {\tilde \epsilon \over 2}}\over  (p_1 - k)^{1 - \tilde \epsilon} } =  \log(K) \times \ \text{finite}
\end{equation}
for which we have (\ref{0log}). The loop integrals for (b),(c) are respectively 
\begin{align}
\int_0^{p_1 - \epsilon} & {k^{ \tilde \epsilon  \over 2} dk \over (p_2 - k)^{1 + {\tilde \epsilon \over 2}}  (p_1 - k)^{1 - \tilde \epsilon} } = 
\nonumber \\ & \hspace{5mm}  {\log(1 +{p_2 - p_1 \over \epsilon} )    [1 + \Op(\tilde \epsilon)] \over (p_2 - p_1)^{1- { \tilde \epsilon \over 2 }}}  + \text{finite} 
\hspace{20mm} u\equiv p_2-p_1
\\ 
\int_{p_1 -p_2 +  \epsilon}^1 & { dk \over k^{1 + {\tilde \epsilon \over 2} } (k+ p_2 - p_1)^{1-\tilde \epsilon }  (k+ p_2)^{\tilde \epsilon \over 2}  } =  
\nonumber \\ & \hspace{5mm}    {\log(1 + {p_1 - p_2 \over \epsilon})  [1 + \Op(\tilde \epsilon)] \over  (p_1 - p_2)^{1- { \tilde \epsilon  \over 2}} }    +\text{finite}
\hspace{20mm} u\equiv p_1-p_2
\end{align}
Both of them give (\ref{rootreghalf})
\end{itemize}

%%%%%%%%%%%%%%%%%
%%%%%%%%%%%%%%%%%%%%%%%%%%%%%%%%%%%%%%%%%%%%%%%%%%%%%%%%%%%%%%%

%%%%%%%%%%%%%%%%%
%%%%%%%%%%%%%%%%%%%%%%%%%%%%%%%%%%%%%%%%%%%%%%%%%%%%%%%%%%%%%%%

%%%%%%%%%%%%%%%%%%%%%%%%%%%%%%%%%%%%%%%%%%%%%%%%%%%%%
\section{Summary and future directions}
%%%%%%%%%%%%%%%%%%%%%%%%%%%%%%%%%%%%%%%%%%%%%%%%%%%%%
In this paper we studied the emergence of a (chiral) strange metal in $1+1$ as the 
low energy sector above a large fermi surface in the fermionic $\psu(1,1)$ sector of ${\cal N}=4$ SYM.
We have shown that, at the two loop level, the constraint of $SU(N)$ gauge invariance 
develops into a full-blown $\frac{SU(N)_N \otimes SU(N)_N}{SU(N)_{2N}}$ gauged coset model. 
This happens since operators annihilated by all the Kac-Moody generators $J^{u}_{n>0}$ 
have zero anomalous dimension when the fermi surface is large, 
up to corrections of the order of the inverse size of the fermi surface, 
and are gapped from the rest of the states. Assuming their conjectured dual $AdS$ states, we obtain a higher spin theory at the near horizon of certain black holes. 

It would be very nice to prove that the chiral strange metal survives 
to all orders in perturbation theory. This would, most likely, require supplementing the large $K$ diagrammatic techniques developed in this paper by the large $K$ limit of the all-order ansatz for 
the dilatation generator given in \cite{Zwiebel:2008gr}. 
Furthermore, one can try to extend this construction to the full $\psu(1,1|2)$ sector. 
If one can find other ground states within this setup, it would be interesting to understand 
what their dual solutions are. These are possibly hairy black holes, or black saturn-like configurations. 

Working out the spectrum of the strange metal at large $N$ limit would also be useful, 
both in guiding the search for an all-order proof, and in using these results within the context of AdS/CFT.  The dual black hole should have, according to this picture, massless higher-spin excitations arising in the near-horizon limit of the singular black hole degeneration, 
which should reproduce the extended W-symmetry of the chiral strange metal. 

A step in this direction would be to compute the quasinormal modes of known SUGRA and stringy fields in the 
bulk (assuming their action can be found reliably) and show that  they match the 
spectrum of the chiral strange metal. While the black hole background is far from simple, 
it is possible that one can apply the methods used to find the quasinormal modes of the dilaton-axion pair in \cite{Micha}. 

\acknowledgments
We would like to thank R.~Gopakumar, M. R.~Gaberdiel, M. Isachenkov for useful discussions. P.N is
grateful for the support by Feinberg fellowship and VATAT fellowship. 

This work was supported in part by an Israel Science Foundation (ISF) center of excellence grant, by the German-Israeli Foundation for Scientific Research and Development and by the Minerva Foundation.
\newpage

%%%%%%%%%%%%%%%%%%%%%%%%%%%%%%%%%%%%%%%%%%%%%%%%%%%%%%%%%%%%%%%%%%%%%%%%%%%
\appendix
%%%%%%%%%%%%%%%%%%%%%%%%%%%%%%%%%%%%%%%%%%%%%%%%%%%%%%%%%%%%%%%%%%%%%%%%%%%

\section*{Appendices}

%%%%%%%%%%%%%%%%%%%%%%%%%%%%%%%%%%%%%%%%%%%%%%%%%%

\section{One loop dilatation in terms of currents}
\label{current_appendix}

%%%%%%%%%%%%%%%%%%%%%%%%%%%%%%%%%%%%%%%%%%%%%%%%%%
\subsection{$\dD_2$: An explicit evaluation}
%%%%%%%%%%%%%%%%%%%%%%%%%%%%%%%%%%%%%%%%%%%%%%%%%%%%%%%%%%%%%%%%%%%%%%%%%%
In this appendix, we obtain the expression for $\dD_2$ given in (\ref{D2inrho}) from the definition of $\dD_2 = 2 \{I^+,\bar I^- \}$. To start with, it is convenient to rewrite $I^+,\bar I^-$ given in (\ref{Iporig},\ref{Imorig}) explicitly in terms of structure constants. 
\begin{eqnarray}
I^+ &=& {i \over 2 \sqrt 2} f^{abc}  \sum_{k,q = 0}^\infty \sqrt{k+q+2 \over (k+1)(q+1)} \  \rho^a_k \rho^b_q   \check \rho^c_{k+q+1} \\
\bar I^- &=& {i \over 2 \sqrt 2 N } f^{def} \sum_{m,n =0}^{\infty} \sqrt{m+n+2 \over (m+1)(n+1)} \rho^d_{m+n+1} \check \rho^e_n \check \rho^f_m 
\end{eqnarray}
Now we compute $\dD_2 = 2 \{I^+,\bar I^- \}$ by anticommuting the fermions. We get
\begin{eqnarray}
- 8 N \{I^+,\bar I^- \} &=& 
\nonumber  
f^{abc} f^{cef} \sum_{k,q= 0} {k+q + 2 \over \sqrt{(m+1)(n+1) (k+1) (q+1) } }  \delta_{k+q, m+n}  \rho^a_k \rho^b_q \check \rho^e_n   \check \rho^f_m\\
\nonumber
&& + 4 f^{abc} f^{dea} \sum_{k,n,q = 0}^\infty {1 \over k+1} \sqrt{(k+q+2)(k+n+2) \over(q+1)(n+1)}  \rho^d_{k+n+1}  \check \rho^e_n \rho^b_q    \check \rho^c_{k+q+1} \\ 
\label{compexp}
&& + 2 f^{abc} f^{abd} \sum_{k,q=0}^{\infty} \frac{k+q+2}{(k+1)(q+1)} \rho^d_{k+q+1} \check\rho^c_{k+q+1}
\end{eqnarray}
Let us simplify each of the three terms of this equation
\begin{itemize}
\item Second line of (\ref{compexp}) can be rewritten as 
\begin{eqnarray}\label{partI}
&& 4 \sum_{k=1}^\infty {1 \over k} \left[ \sum_n \sqrt{k+n+1 \over n+1} f^{ade} \rho^d_{k+n} \check \rho^e_n \right]   \left[  \sum_q  \sqrt{k+q+1 \over q+1} f^{abc} \rho^b_q \check \rho^e_{k+q} \right] 
\end{eqnarray}
\item The last line of (\ref{compexp}) is easily seen to be 
\begin{equation}
8N \sum_{k=0} h(k) \rho_k^a \rho_k^a
\end{equation}
\item At last we consider the first line of (\ref{compexp}). Before we start simplifying this term, it is convenient to define 
\begin{equation}
f_{mnkq} = {k+q + 2 \over \sqrt{(m+1)(n+1) (k+1) (q+1) } }
\end{equation} 
Using this definition and Jacobi Identity for structure constants, we can simplify (\ref{compexp}) as

\begin{eqnarray}
\nonumber
&&  \left( f^{aec} f^{cbf} + f^{afc} f^{ceb} \right) \sum_{mnkq} f_{mnkq} \ \delta_{k+q,m+n} \rho^a_k \rho^b_q \check \rho^e_n   \check \rho^f_m \\
\nonumber
&=&  2  f^{aec} f^{cbf} \sum_{mnkq} \left( \  2 \theta(m-q) + \delta_{q,m} \ \right)  f_{mnkq} \  \delta_{k+q,m+n} \rho^a_k \rho^b_q \check \rho^e_n    \check \rho^f_m \\
\nonumber
&=& - 4  \sum_{mnkq} \theta(m-q) f_{mnkq} \delta_{k+q, m+n} f^{aec} \rho^a_k \check \rho^e_n f^{cbf}  \rho^b_q  \check \rho^f_m   + 4 \sum_{q,k} \theta(k-q) f_{kqkq} f^{aec}f^{cef} \rho^a_k \check \rho^f_k \\
\label{dd2exp}
&& \hspace{10mm} + 2 f^{aec} f^{cbf} \sum_{k,q=0} f_{qkkq} \rho^a_k \rho^b_q \check \rho^e_k \check \rho^f_q 
\end{eqnarray}
We analyze each one of the above terms. First term of (\ref{dd2exp}) gives
\begin{eqnarray}
&&-4 \sum_{n,q =0 , \tilde m =1}  f_{\tilde m+q, n, m n + \tilde m, q}  f^{aec} \rho^a_{\tilde m + n} \check \rho^e_n f^{cbf}  \rho^b_q  \check \rho^f_{\tilde m + q}   \\ \nonumber
&=& -4 \sum_{n,q=0, m= 1} {n + m+ q +2 \over \sqrt{ (m+q+1) (n+1)(n+m+1)(q+1) } } f^{aec}  \rho^a_{  m + n} \check \rho^e_n f^{cbf}  \rho^b_q  \check \rho^f_{ m + q}
\end{eqnarray}
Note that this combined with (\ref{partI}) gives
\begin{eqnarray}\nonumber
&& \sum_{n,q=0, k=1} {1 \over \sqrt{(q+1) (n+1)} } \left[ \sqrt{(k+n+1)(k+q+1) \over k} - {k+ n+ q +2  \over \sqrt{(k+n+1) (k+q+1)  }} \right] \\ \nonumber
&& \hspace{10mm} \left[ f^{ade} \rho^d_{k+n} \check \rho^e_n \right]   \left[   f^{abc} \rho^b_q \check \rho^e_{k+q} \right]  \\
&=& 4 \sum_{n,q=0, k=1} \frac{\sqrt{(n+1) (q+1)}}{k \sqrt{(k+n+1) (k+q+1)}}  \left[ f^{ade} \rho^d_{k+n} \check \rho^e_n \right]   \left[   f^{abc} \rho^b_q \check \rho^e_{k+q} \right]
\end{eqnarray}
which matches with the four fermion term in (\ref{D2inrho}).  Meanwhile second term of (\ref{dd2exp}) gives
\begin{eqnarray}
&&4 \sum_{k=0}^\infty \sum_{q= 0}^{k-1} {k+q + 2 \over (k+1)  (q+1)} (-2 N) \rho^a_k \check \rho^a_k 
%\nonumber \\ \nonumber
%&=& - 8 N  \sum_{k=0}^\infty  \rho^a_k \check \rho^a_k \sum_{q= 0}^{k-1} \left( {1 \over k+1 } + {1 \over q+1 } \right) \\ 
%&=& 
-8 N \sum_{k=0}^\infty  \rho^a_k \check \rho^a_k \left( h(k) + {k \over k+1} \right)
\end{eqnarray}
It is also useful to use the identity 
\begin{equation}
\rho^a_k \check \rho^a_k = {1 \over 2 N} f^{aec} f^{fbc} \sum_{q=0}^\infty \rho^a_k \rho^b_q \check  \rho^e_k \check \rho^f_q
\end{equation}
The last term of (\ref{dd2exp}) gives 
\begin{align}
\nonumber
  2  f^{aec}  f^{cbf} & \sum_{k,q} {k+q+2 \over (k+1)(q+1)}  \rho^a_k \rho^b_q \check \rho^e_k \check \rho^f_q  
= \sum_k { 4 \over k+1 } f^{aec} f^{cbf} \sum_q \rho^a_k \rho^b_q \check  \rho^e_k \check \rho^f_q \\ & =- \sum_k { 8 N \over k+1 }  \rho^a_k \tilde \rho^a_k
\end{align}
Summing all these contributions yields  (\ref{D2inrho}).
\end{itemize}

%%%%%%%%%%%%%%%%%%%%%%%%%%%%%%%%%%%%%%%%%%%%%%%%%%
\subsection{Current Algebra}
\label{current_algebra_appendix} 
%%%%%%%%%%%%%%%%%%%%%%%%%%%%%%%%%%%%%%%%%%%%%%%%%%
In this Appendix, we compute the commutation relations of the $J^a_n$ to show that under specific limits, it reproduces (\ref{kacmoody}) of the current algebra. Consider $[J^a_p,J^b_q]$ for $p,q>0$.
\begin{eqnarray}
[J^a_p,J^b_q] %&=& - f^{acd} f^{bgh} \sum_{m,n=0} \sqrt{(m+1)(n+1) \over (m+p+1)(n+q+1)} [\rho^c_m \check \rho^d_{m+p}, \rho^g_n \check \rho^h_{n+q}]  \nonumber \\
%&=&  -  f^{acd} f^{bgh} \sum_{m,n=0}  \sqrt{(m+1)(n+1) \over (m+p+1)(n+q+1)}  \left( \delta^{dg} \delta_{m+p,n} \rho^c_m \check \rho^h_{n+q} - \delta^{hc} \delta_{m,n+q} \rho^g_n \check \rho^d_{m+p}  \right) \nonumber  \\
&=& -\left( f^{acd} f^{bdh} - a \leftrightarrow b \right) \sum_n \sqrt{n+1 \over n+p+q+1} \rho^c_n \check \rho^h_{n+p+q} \nonumber \\
%&=& f^{abd} f^{dch} \sum_n  \sqrt{n+1 \over n+p+q+1} \rho^c_n \check \rho^h_{n+p+q} \nonumber \\
&=& i f^{abc} J^c_{p+q},
\end{eqnarray}
where in the last step we used the Jacobi identity.
Similarly $[ J^a_{-p}, J^b_{-q} ] = i f^{abc} J^c_{-p-q}$. 
Now, consider $[J^a_p,J^d_{-q}]$ for $p \geq q$ and $q>0$
\begin{eqnarray}
[J^a_p,J^d_{-q}] &=& - f^{abc} f^{def} \sum_{m,n=0}  g(m,n)   [\rho^b_m \check \rho^c_{m+p}, \rho^e_{n+q} \check \rho^f_{n}]  \\
%&=&    \sum_{m,n=0} g(m,n) \left(-  \delta_{n,m+p-q} f^{abc} f^{dcf}  \rho^b_m \check \rho^f_{n} +  \delta_{m,n} f^{dbc} f^{acf}  \rho^b_{n+q} \check \rho^f_{m+p}  \right) \nonumber \\
&=& \sum_{m=q}^{\infty} g(m-q,m-q) f^{dbc} f^{acf} \rho^b_m \check \rho^f_{m+p-q}   - \sum_{m=0}^\infty f^{abc} f^{dcf} g(m,m+p-q) \rho^b_m \check \rho^f_{m+p-q} \nonumber \\
%&=& \left( \sum_{m=0}^{\infty} {m-q+1 \over \sqrt{(m+p-q+1)(m+1)}} + \sum_{m=0}^{q-1}  {q-1-m \over \sqrt{(m+p-q+1)(m+1)}}  \right) f^{dbc} f^{acf} \rho^b_m \check \rho^f_{m+p-q}   \nonumber \\
%&& \hspace{20mm} - \sum_{m=0}^\infty  f^{abc} f^{dcf}  { \sqrt{(m+1)(m+p-q+1)} \over m+p+1 }\rho^b_m \check \rho^f_{m+p-q} 	  \nonumber \\
&=& i f^{adc} J^c_{p-q} +  \sum_{m=0}^{\infty} \sqrt{ m+1  \over m+p-q+1}  \left({ q f^{abc} f^{dcf} \over m+p+1} - {q f^{dbc} f^{acf}  \over m+1 } \right)   \rho^b_m \check \rho^f_{m+p-q}  \nonumber   \\
&& \hspace{20mm} +  f^{dbc} f^{acf}   \sum_{m=0}^{q-1} {q-1-m \over \sqrt{(m+p-q+1)(m+1)} }  	\rho^b_m \check \rho^f_{m+p-q} \nonumber \\ 
&=& i f^{adc}  J^c_{p-q} -  \sum_{m=0}^{\infty} \sqrt{ m+1  \over m+p-q+1}  \left({ q f^{abc} f^{dcf} \over m+p+1} - {q f^{dbc} f^{acf}  \over m+1 } \right)   \check \rho^f_{m+p-q} \rho^b_m  \nonumber   \\
&& \hspace{20mm} -  f^{dbc} f^{acf}   \sum_{m=0}^{q-1} {q-1-m \over \sqrt{(m+p-q+1)(m+1)} }  \check \rho^f_{m+p-q}	\rho^b_m  \nonumber \\ 
&& \hspace{20mm} -  2N\delta_{p,q} \delta^{ad} \left\lbrace \sum_{m=0}^\infty   \left[ {q \over m+ q +1} - {q \over m+1} \right]  + \sum_{m=0}^{q-1} {q - 1 -m \over m+1} \right\rbrace \nonumber 
\end{eqnarray}
where $g(m,n) = \sqrt{(m+1)(n+1) \over(m+q+1)(n+p+1)}$. If we now act on fluctuation states $\ket F$ defined previously, following simplification occurs. The terms proportional to $\check \rho \rho$ above are nonzero only if $K+s+q-p\ge m \ge K-s$, i.e $m\sim K$ and hence these two terms are $\Op(1/K)$ and can be dropped. Then, we have
\begin{eqnarray}
[J^a_p,J^d_{-q}] &=& i f^{adc} J^c_{p-q} + (2 N q) \delta^{ab} \delta_{pq} + \Op(1/K)
\end{eqnarray}
One can perform a similar computation with $p<q$. The result can be summarized (again upto \Op(1/K) corrections)
\begin{equation}
[J^a_p,J^d_{-q}] = i f^{adc} J^c_{p-q} + (2 N q) \delta^{ab} \delta_{pq}
\end{equation}
To summarize, we can write
\begin{equation}
[J^a_m,J^b_{n}]  =  i f^{abc} J^c_{m+n}  + 2Nm \ \delta_{m+n,0}\ \delta^{ab} 
\end{equation}

%%%%%%%%%%%%%%%%%%%%%%%%%%%%%%%%%%%%%%%%%%%%%%%%%%%%%
\subsection{Computing $\dD_2$ on generic fluctuations}
\label{appendix_d2_fluct}
%%%%%%%%%%%%%%%%%%%%%%%%%%%%%%%%%%%%%%%%%%%%%%%%%%%%%
In this appendix we work out $\dD_2$ for a particle-hole state in the continuum limit.
This computation can be easily carried out without taking the continuum limit, but we shall use it as an example of the formalism. 
The computation shows that a typical fluctuation would have $\Op(1)$ value for $\dD_2$ (after subtracting the contribution of the number operator). 

First define the operator $\ket {\Op_F} \equiv C \int dx dy f(x,y) \rho^a(1+x) \check \rho^a(1-y) \ket {\Op^{(K)} }$. The function $f(x,y)$ satisfies $\int dx dy |f(x,y)|^2 = 1$ and $C= N^2-1$. Note that for the state to be a small fluctuation, $f(x,y)$ is nonvanishing only if $x,y \in (1 - \delta, 1+ \delta)$.

The two fermion part of $\dD_2$ does not contribute to the difference  $\bra {\Op_F}   \dD_2 \ket {\Op_F} -  \bra{ \Op^{(K)} } \dD_2 \ket {\Op^{(K)}} $. The only remaining term is
\begin{equation}
{1 \over N} \int {dz  \over z} |J^a(z) \Op(x,y)|^2
\end{equation}
One can show that 
\begin{eqnarray}\nonumber
J^a(z)  \ket{\Op_F } &=&  - i f^{abc} \int dx dy f(x,y) \left( \sqrt{1 + x-z \over 1 + x} \rho^b(1+x-z) \check \rho^c(1-y) \right. \\
&& \hspace{20mm} \left. - \sqrt{1 -y \over 1-y +z }\rho^b(1+x) \check \rho^c(1+z-y)    \right) 
\end{eqnarray}
Since $f(x,y)$ has support only in a range $(1 - \delta,1 + \delta)$, the momenta factor can be simplified to
\begin{align}
J^a(z) & \ket{\Op_F } = - i f^{abc} \int dx dy f(x,y) \Big[   \rho^b(1+x-z) \check \rho^c(1-y) - \nonumber \\ 
& \hspace{20mm}\rho^b(1+x) \check \rho^c(1+z-y)    \Big] \ket{\Op^{(K)}} +  \Op(1/K)
\end{align}
Again the first of the above term is nonzero only if $z\le x$ while the second is nonzero only if $z\le y$. A similar computation as in the discrete case gives
\begin{eqnarray} 
{1 \over N} \int {dz  \over z} |J^a(z) \Op(x,y)|^2 = 2 \int dx dy |f(x,y)|^2 \left[ \log(x/\epsilon) +  \log(y/\epsilon)  \right] 
\end{eqnarray}
If $f(x,y)$ is localized around $x_0,y_0$, then we can  further simplify this to 
\begin{equation}
{1 \over N} \int {dz  \over z} |J^a(z) \Op(x,y)|^2 = 2 \left[ \log(x_0/\epsilon) +  \log(y_0/\epsilon)   \right]
\end{equation}
Since $x_0 /\epsilon$ is finite, this is a $\Op(1)$ quantity. 

\section{Simplifying $\dD_4$} \label{d4details}

%%%%%%%%%%%%%%%%%%%%%%%%%%%%%%%%%%%%%%%%%%%%%%%%%%%%
In this appendix we will derive the expressions for $V,C$ as given in (\ref{Vexp},\ref{Cexp}) from their definitions given (\ref{Vdef},\ref{Cdef}).  For completeness, we give below the expressions for all $I^\pm, \bar{I}^\pm$ to leading order in $g$ (as given in (3.28) of \cite{Dori}) 
\begin{subequations}\label{supercharges}
\begin{align}
    I^+_1
    =&\frac1{\sqrt2}\sum_{k,q=0}^\infty
    \biggl(\sqrt{\frac{k+q+2}{(k+1)(q+1)}}\tr\cno{\rho_{(k)}\rho_{(q)}\check{\rho}_{(k+q+1)}}
    +\sum_{i=2}^3\tr\cno{\frac{1}{\sqrt{k+1}}
    \left[\rho_{(k)},\phi^i_{(q)}\right]\check{\phi}^i_{(k+q+1)}} 
     \nonumber \\ 
    &\hspace{4em}
    +\sqrt{\frac{q+1}{(k+1)(k+q+2)}}\tr\cno{\left\{\rho_{(k)},\bar\rho_{(q)}\right\}\check{\bar\rho}_{(k+q+1)}}
     \nonumber \\ &\hspace{4em}
    -\frac{1}{\sqrt{k+q+1}}\tr\cno{\left[\phi^2_{(k)},\phi^3_{(q)}\right]\check{\bar\rho}_{(k+q)}}
    \biggr) \\
    \bar I^+_1
    =&\frac1{\sqrt2}\sum_{k,q=0}^\infty\biggl(\sqrt{\frac{k+q+2}{(k+1)(q+1)}}
    \tr\cno{\bar\rho_{(k)}\bar\rho_{(q)}\check{\bar\rho}_{(k+q+1)}}
    +\sum_{i=2}^3\frac{1}{\sqrt{k+1}}\tr\cno{\left[\bar\rho_{(k)},\phi^i_{(q)}\right]\check{\phi}^i_{(k+q+1)}}
    \nonumber \\ &\hspace{4em}
    +\sqrt{\frac{q+1}{(k+q+2)(k+1)}}\tr\cno{\left\{\bar\rho_{(k)},\rho_{(q)}\right\}\check{\rho}_{(k+q+1)}}
    \nonumber \\ &\hspace{4em}
    +\frac1{\sqrt{k+q+1}}\tr\cno{\left[\phi^2_{(k)},\phi^3_{(q)}\right]\check{\rho}_{(k+q)}}
    \biggr)
\\
    I^-_1
    =&\frac1{\sqrt2N}\sum_{m,n=0}^\infty\biggl(
    \sqrt{\frac{n+m+2}{(n+1)(m+1)}}\tr\cno{\bar\rho_{(n+m+1)}\check{\bar\rho}_{(n)}\check{\bar\rho}_{(m)}}
    -
    \frac{1}{\sqrt{n+m+1}}\tr\cno{\rho_{(n+m)}\left[\check{\phi}^2_{(n)},\check{\phi}^3_{(m)}\right]}
    \nonumber \\ & \hspace{4em}+\sqrt{\frac{m+1}{(n+m+2)(n+1)}}    \tr\cno{\rho_{(n+m+1)}\left\{\check{\bar\rho}_{(n)},\check{\rho}_{(m)}\right\}}
    \nonumber \\ &  \hspace{4em}
    +\frac1{\sqrt{n+1}}\sum_{i=2}^3
    \tr\cno{\phi^i_{(n+m+1)}\left[\check{\bar\rho}_{(n)},\check{\phi}^i_{(m)}\right]}
    \biggr)\\    
    \bar I^-_1
        =&\frac{1}{\sqrt2N}\sum_{m,n=0}^\infty\biggl(
        \sqrt{\frac{n+m+2}{(n+1)(m+1)}}
        \tr\cno{\rho_{(n+m+1)}\check{\rho}_{(n)}\check{\rho}_{(m)}}
        +
       	\frac{1}{\sqrt{n+m+1}}\tr\cno{\bar\rho_{(n+m)}\left[\check{\phi}^2_{(n)},\check{\phi}^3_{(m)}\right]}
         \nonumber \\ &\hspace{4em}
        +\sqrt{\frac{m+1}{(n+1)(n+m+2)}}
        \tr\cno{\bar \rho_{(n+m+1)}\left\{\check{\rho}_{(n)},\check{\bar \rho}_{(m)}\right\}}
        \nonumber \\ &\hspace{4em}
        +\frac1{\sqrt{n+1}}\sum_{i=2}^3
        \tr\cno{\phi^i_{(n+m+1)}\left[\check{\rho}_{(n)},\check{\phi}^i_{(m)}\right]}
        \biggr)~.
\end{align}
\end{subequations}

Complexity of the calculation is reduced vastly if we use the fact that we are interested only in states of the form $\bra f \dD_4 \ket f$ where $\ket f$ belongs to the fermionic $\su(1,1)$ sector.  To see this 
\begin{itemize}
\item Note that the definition  (\ref{expd4}) of $\dD_4$ involves $I^+,C,\bar I^-$. Since the action of $I^+, \bar I^-$ closes on fermionic  $\su(1,1)$ sector, the only relevant matrix element we need to compute are of the form $\bra f C \ket f$ and $\bra f V  \ket f$. 
\item Note that the definition (\ref{Vdef},\ref{Cdef}) $V,C$ involve $I^-,\bar I^+,h$. Since $I^-$ ($\bar I^+$) acting on right (left) on state $ \ket f$ vanishes, the only ordering which survives in $\bra f C \ket f$ and $\bra f V  \ket f$ is one where $I^-$ occurs to left of $\bar I^+$. Now since commutation with $h$ leads to same parton structure (albeit with different coefficients), it is easy to see that we only need to keep terms in $I^-$ ($\bar I^+$) which are nonvanishing when acting on right (left)  on state $ \ket f$.
\end{itemize}
Thus it is enough to restrict to the following terms in $I^-,\bar I^+$
\begin{eqnarray}     \nonumber
\sqrt{2N} I^-_1
    &=&  i f^{def} \sum_{m,n=0}^\infty  \left[ \sqrt{\frac{m+1}{(n+m+2)(n+1)}}    \rho^d_{n+m+1} \check{\bar\rho}^e_n \check{\rho}^f_m      \right. \\
 && \hspace{20mm} 
   \left. -\frac{1}{\sqrt{n+m+1}} \rho^a_{n+m} \check{\phi}^{2b}_n \check{\phi}^{3c}_m   \right]  \\
    \nonumber    
\sqrt 2 \bar I^+_1 
    &=&  i f^{abc }\sum_{k,q=0}^\infty \left[ \sqrt{\frac{q+1}{(k+q+2)(k+1)}}  \bar\rho^a_k \rho^b_q  \check{\rho}^c_{k+q+1}  \right. \\
    && \hspace{20mm}
   \left. +\frac1{\sqrt{k+q+1}} \phi^{2a}_k  \phi^{3b}_q \check{\rho}^c_{k+q}  \right] \\
   h &=& \sum_{n=0}^\infty {h(n+1) \over 2} \left[ \rho^a_n \check \rho^a_n +  \bar \rho^a_n \check {\bar\rho}^a_n \right] + \sum_{n=0}^\infty {h(n) \over 2} \sum_{i=2}^3 \phi^{ia}_n \check \phi^{ia}_n
\end{eqnarray}
where we have opened the trace to show the gauge indices explicitly.  Now that we have simplified various supercharges, we  can evaluate the commutators in (\ref{Vdef},\ref{Cdef}) keeping in mind that we are interested only in those terms in $C,V$ which are nonvanishing on the states of fermionic $\su(1,1)$ sector. After some straight forward algebra we get
\begin{eqnarray}
V &=& {i \over 2 N} \sum_{m=0 , u =1}^\infty B_{m,u} \ f^{abc} \rho^b_{m+u} \check \rho^c_m \ J^a_u +  \sum_{m=0}^\infty B_m  \rho^a_m \check \rho^a_m \\
C &=& {i \over 2 N} \sum_{q =0 , u =1}^\infty B_{q,u} \ J^a_{-u} \ f^{abc}  \rho^b_q \check \rho^c_{q+u} +  \sum_{m=0}^\infty B_m  \rho^a_m \check \rho^a_m
\end{eqnarray}
where $B_{m,u} = \sqrt{m+1\over  m+u+1 } \  {h(m+u+1) - h(m+1) -   h(u)  \over u }$ and $B_m  =  h(m+1) - 2$ and the current  $J$  is as defined in (\ref{defJ}).

\section{Two fermion Diagrams in $\dD_4$ between Light States}
\label{appendix_D4_2fermion}
In this appendix we discuss two-fermion terms, which are always of the form 
\begin{equation}
\Op_{2f} = \int dp H(p) \rho^a(p) \check \rho^a(p)
\end{equation}
Since a finite term in $H(p)$ can be shifted away by a chemical potential for $U(1)$, 
we do not need to compute this contribution explicitly. 
However, for completeness we work it out explicitly, finding that the result is $\Op(K^0)$.

\begin{figure}[h!]
\centering
\begin{subfigure}{.4\textwidth}
\centering
\includegraphics[width=.9\textwidth, keepaspectratio]{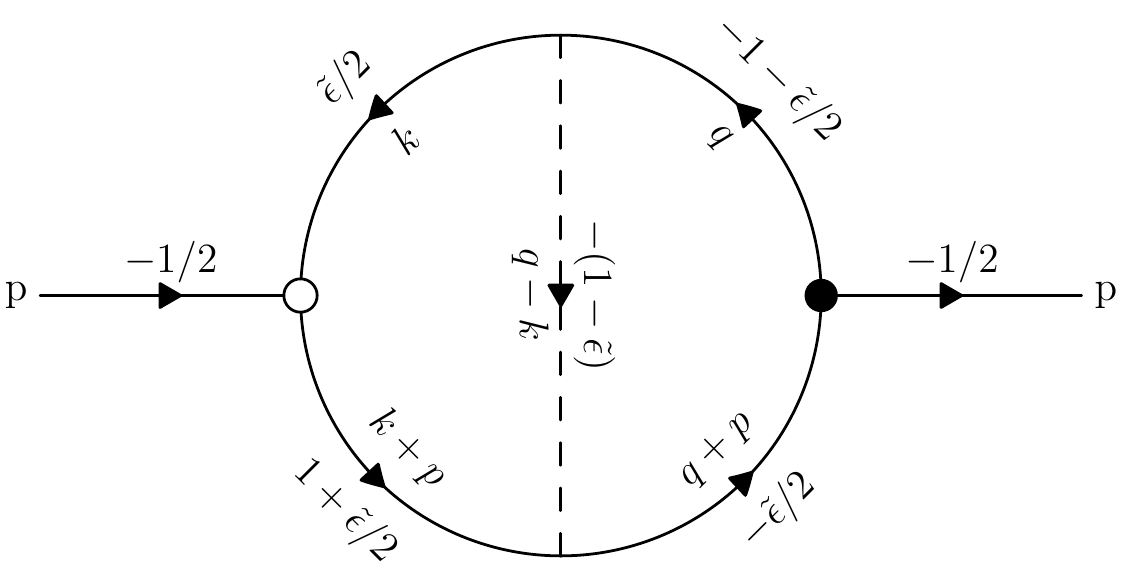}
  \caption{}
\end{subfigure} 
\begin{subfigure}{.4\textwidth}
\centering
  \includegraphics[width=.9\textwidth, keepaspectratio]{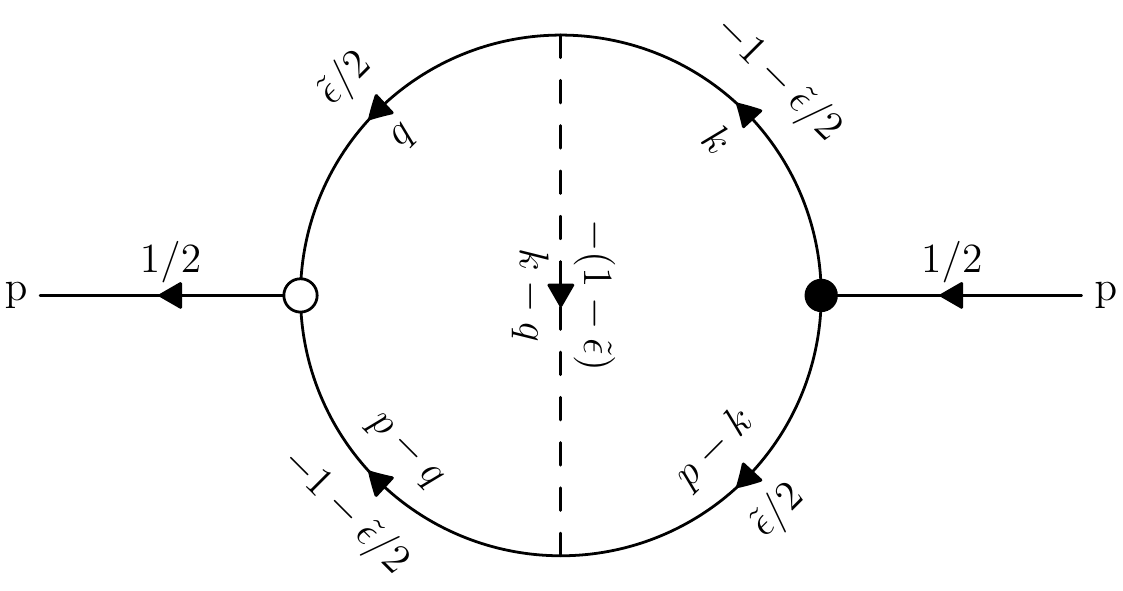}
    \caption{}
\end{subfigure} \\ 
\begin{subfigure}{.4\textwidth}
\centering
  \includegraphics[width=.9\textwidth, keepaspectratio]{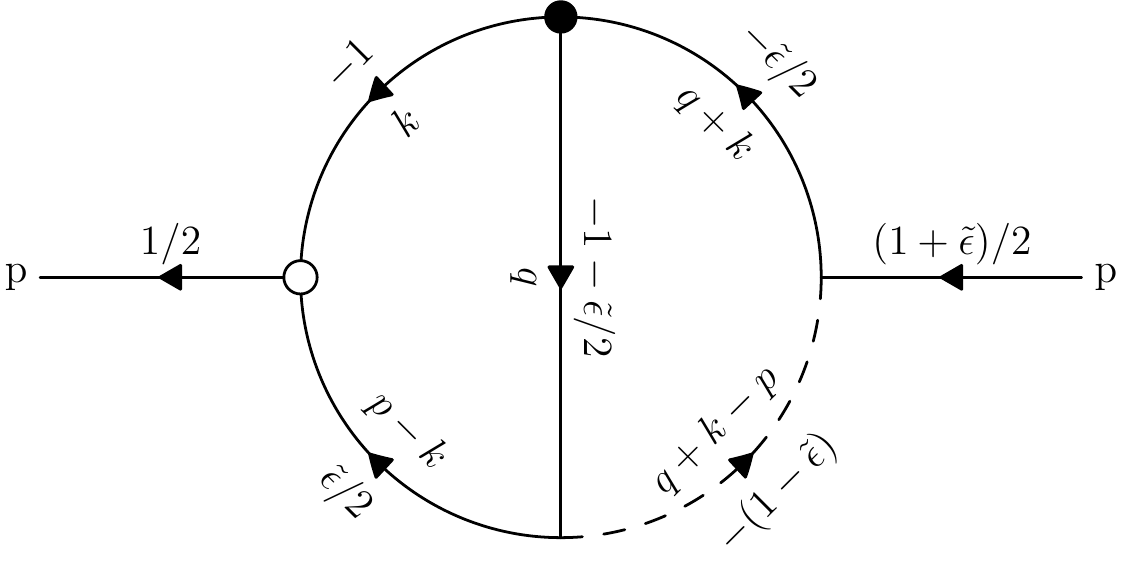}
  \caption{}
\end{subfigure} 
\begin{subfigure}{.4\textwidth}
\centering
  \includegraphics[width=.9\textwidth, keepaspectratio]{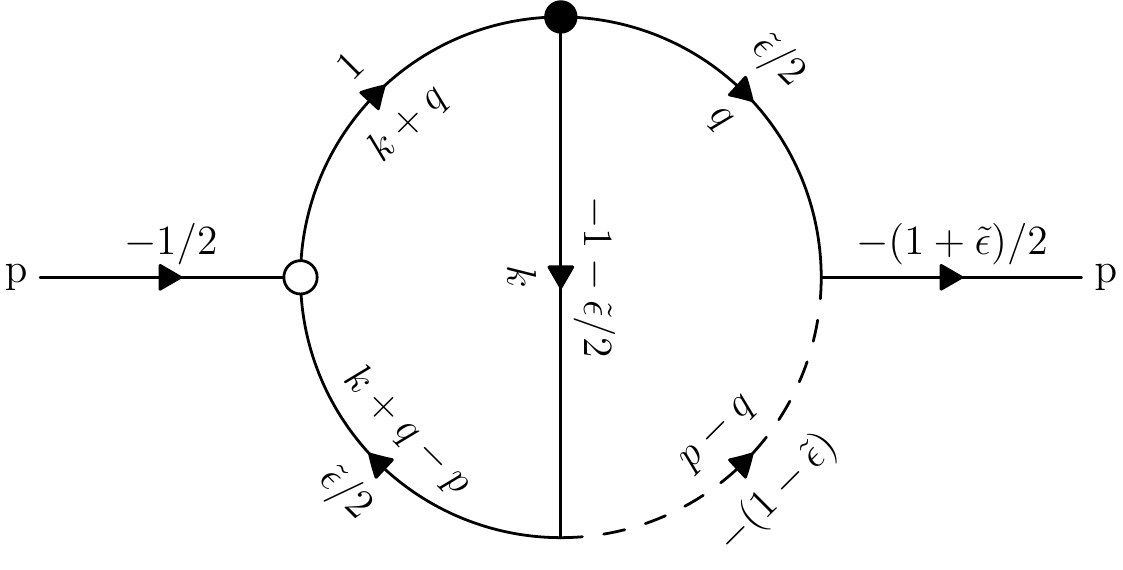}
  \caption{}
\end{subfigure}\\ 
\begin{subfigure}{.4\textwidth}
\centering
  \includegraphics[width=.9\textwidth, keepaspectratio]{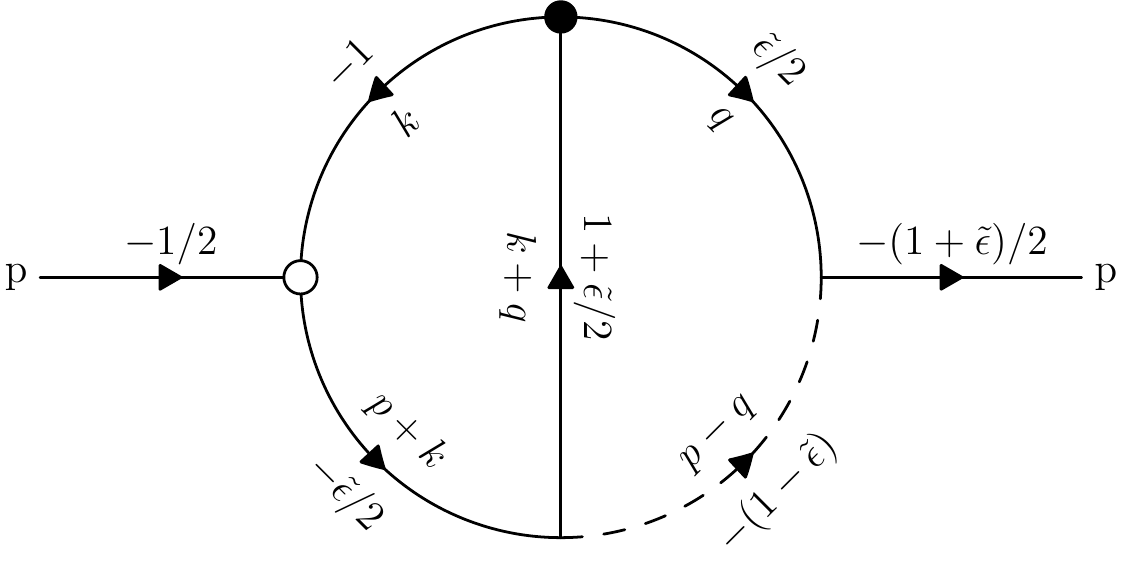}
  \caption{}
\end{subfigure}\\
\caption{1PI corrections to the propagator from $\dD_4$.}
\label{D4F21PI}
\end{figure}

The only possible non-1PI diagram is given  below
\begin{figure}[h!]
\centering
\includegraphics[width=.6\textwidth, keepaspectratio]{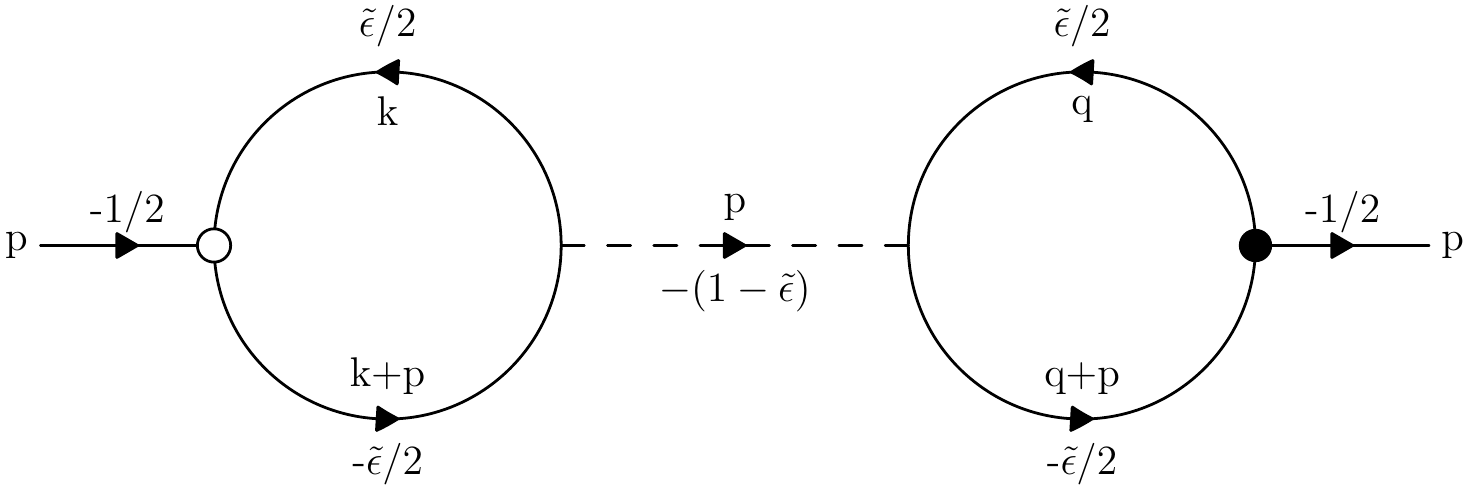}
\caption{Diagram correcting propagator with two disjoint loops.}
\end{figure}
which just gives $H(p) \sim {1 \over p^2}$. The 1PI diagrams are more nontrivial. They are shown in figure \ref{D4F21PI}
which gives loop contribution\footnote{Recall ${\delta \over\epsilon}$ is finite} (in order a,b,c,d,e) 
\begin{align}
(a) &= \int_\epsilon^{1+\delta} {dq \over   q^{1 +{\tilde \epsilon \over 2}} (q+p)^{\tilde \epsilon \over 2}  } \int_\epsilon^{q-\epsilon} dk {(k+p)^{1 +{\tilde \epsilon \over 2}} k^{\tilde \epsilon \over 2} \over   (q-k)^{1 - \tilde \epsilon  }    }\\ % &\sim &  p \log(K)^2 + \log(K) + finite \\
(b) &=  \int_0^1 { dq  \ q^{\tilde \epsilon \over 2} \over  (p-q)^{1 +{\tilde \epsilon \over 2}}} \int_q^p dk {(p-k)^{\tilde \epsilon \over 2}   \over k^{1 +{\tilde \epsilon \over 2}} (k-q)^{1 - \tilde \epsilon  }  } \\  % &\sim &  {1 \over p} \log(K)^2 +  finite
(c) &=  \int_0^1 { dq \over  q^{1 +{\tilde \epsilon \over 2}} } \int_{1-q}^1 dk {(p-k)^{\tilde \epsilon \over 2} \over k \ (q+k-p)^{1 - \tilde \epsilon  } (q+k)^{\tilde \epsilon \over 2} } \\ % &\sim &{1 \over p }  \log(K)^2 +  finite \\
(d) &= \int_0^1 {dq \ q^{\tilde \epsilon \over 2}\over (p-q)^{1 - \tilde \epsilon  }}  \int_{1-q}^1 dk {(k+q)(k+q-p)^{\tilde \epsilon \over 2} \over k^{1 +{\tilde \epsilon \over 2}} } \\ % &\sim &  p \log(K)^2 + p \log(K) + finite  \\
(e) &= \int_0^1 {dq \ q^{\tilde \epsilon \over 2}\over (p-q)^{1 - \tilde \epsilon  }}  \int_{1-q}^1 dk {(k+q)^{1 +{\tilde \epsilon \over 2}}    \over k  (p+k)^{\tilde \epsilon \over 2} } % &\sim &  p \log(K)^2 + p \log(K) + finite   
\end{align}
Diagram (b) is its own conjugate whereas all the other diagrams have a hermitian conjugate diagram which we have not written.  

These integrals can be evaluated, and yield, in the large $K$ limit, results of the form
\begin{equation}
\label{logform}
c_1 \log^{2}K + c_2 \log K + c_3. 
\end{equation}

\begin{figure}[h!]
\centering
\begin{subfigure}{.4\textwidth}
\centering
\includegraphics[width=.9\textwidth, keepaspectratio]{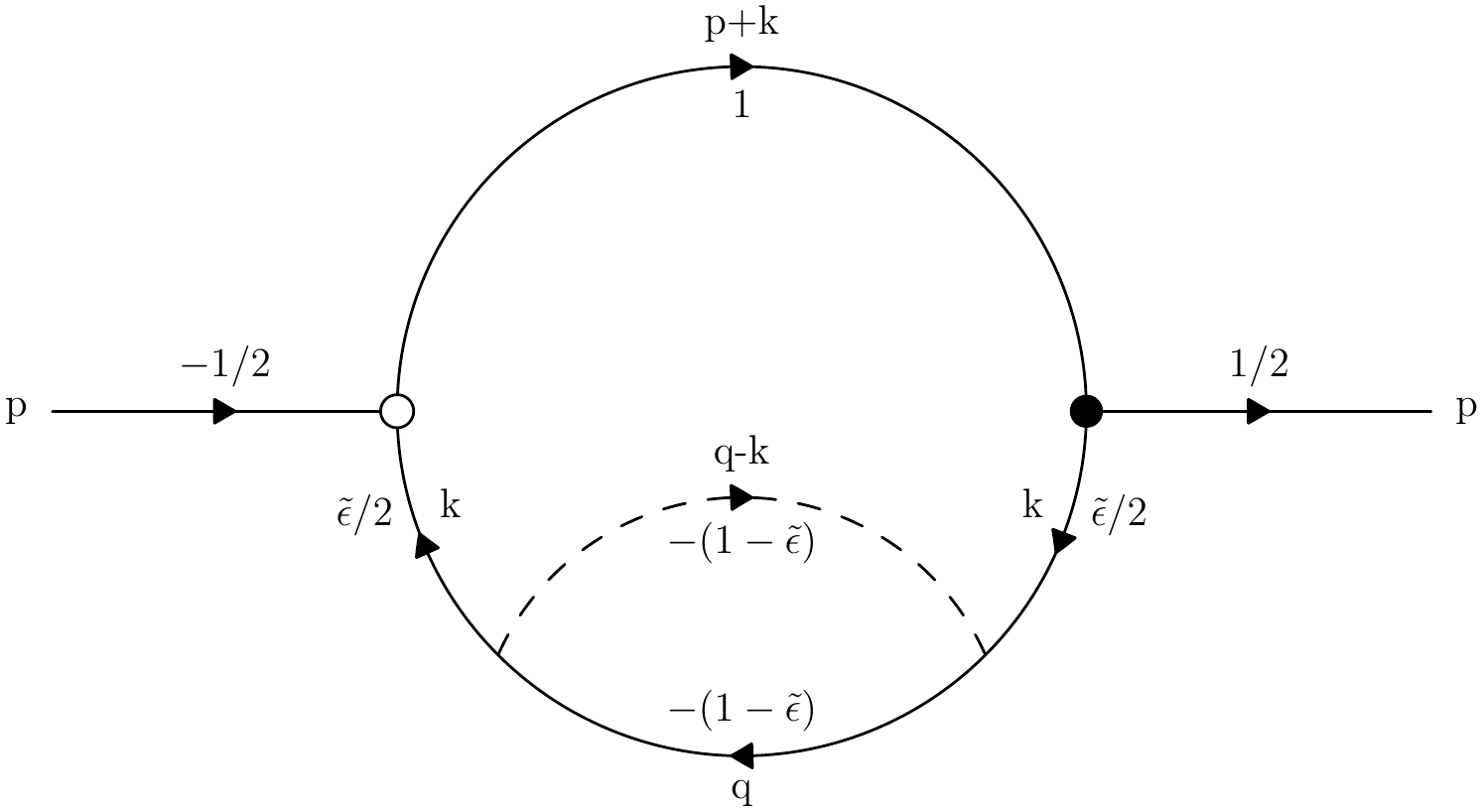}
  \caption{}
\end{subfigure} 
\begin{subfigure}{.4\textwidth}
\centering
  \includegraphics[width=.9\textwidth, keepaspectratio]{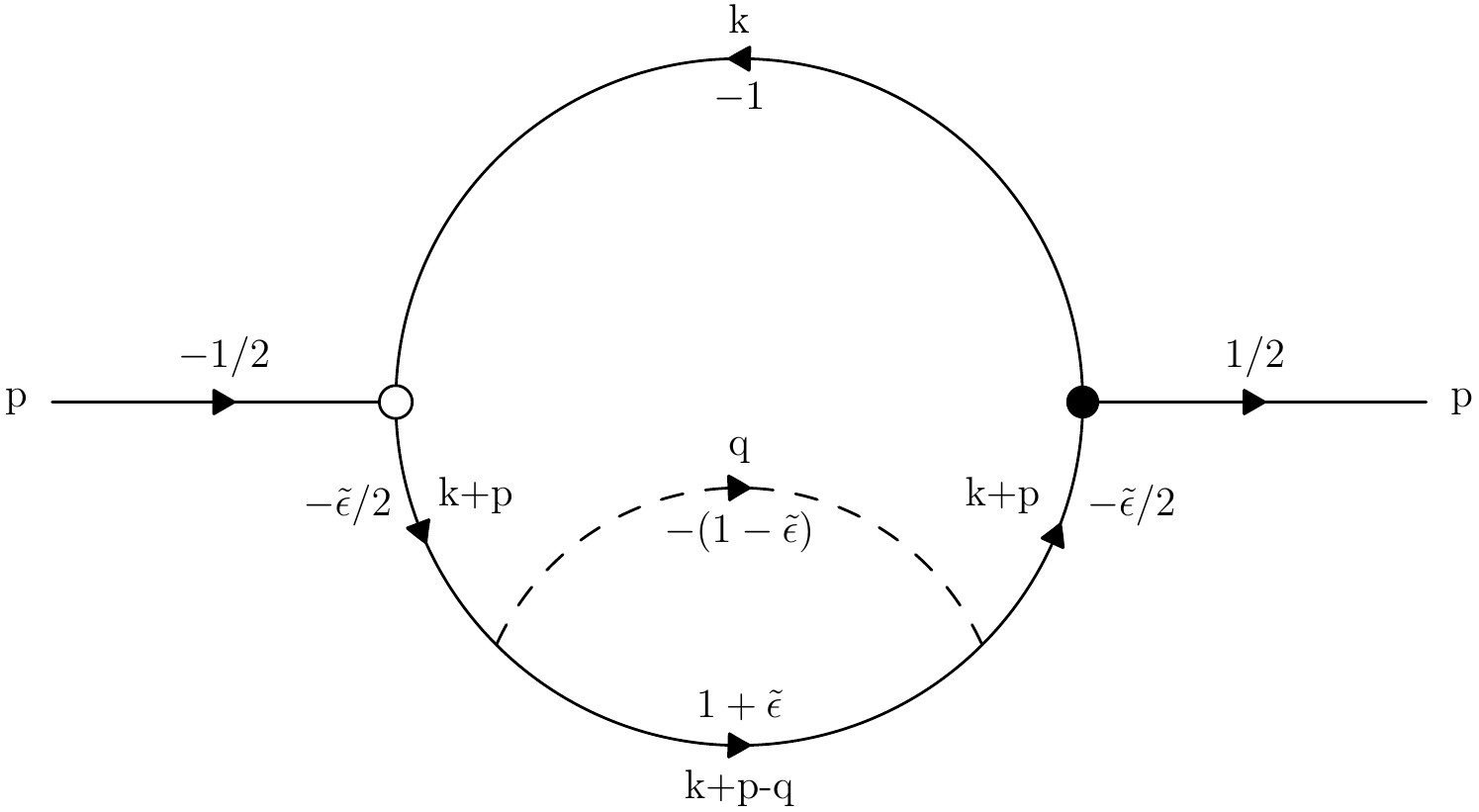}
    \caption{}
\end{subfigure} \\ 
\begin{subfigure}{.4\textwidth}
\centering
  \includegraphics[width=.9\textwidth, keepaspectratio]{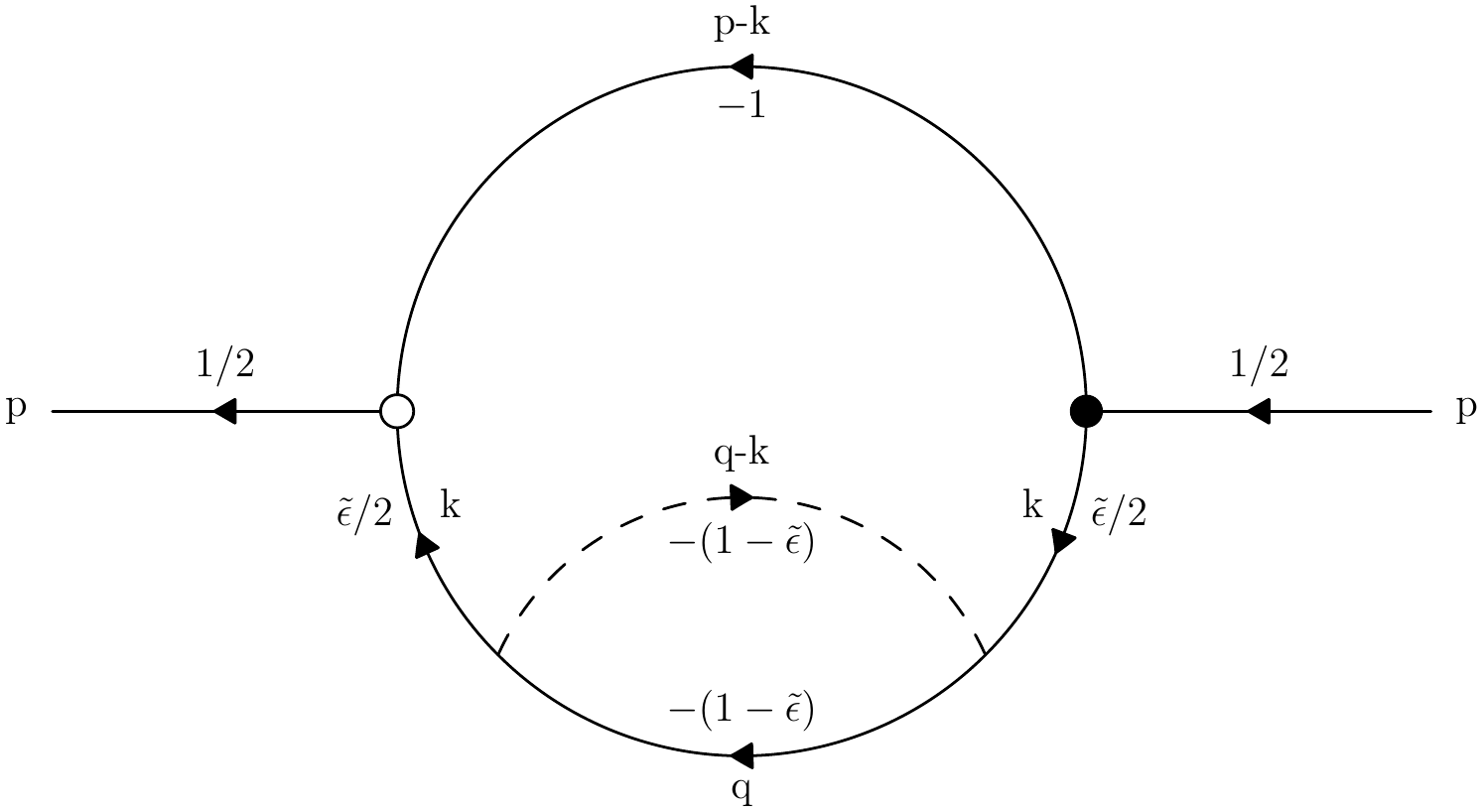}
  \caption{}
\end{subfigure} 
\begin{subfigure}{.4\textwidth}
\centering
  \includegraphics[width=.9\textwidth, keepaspectratio]{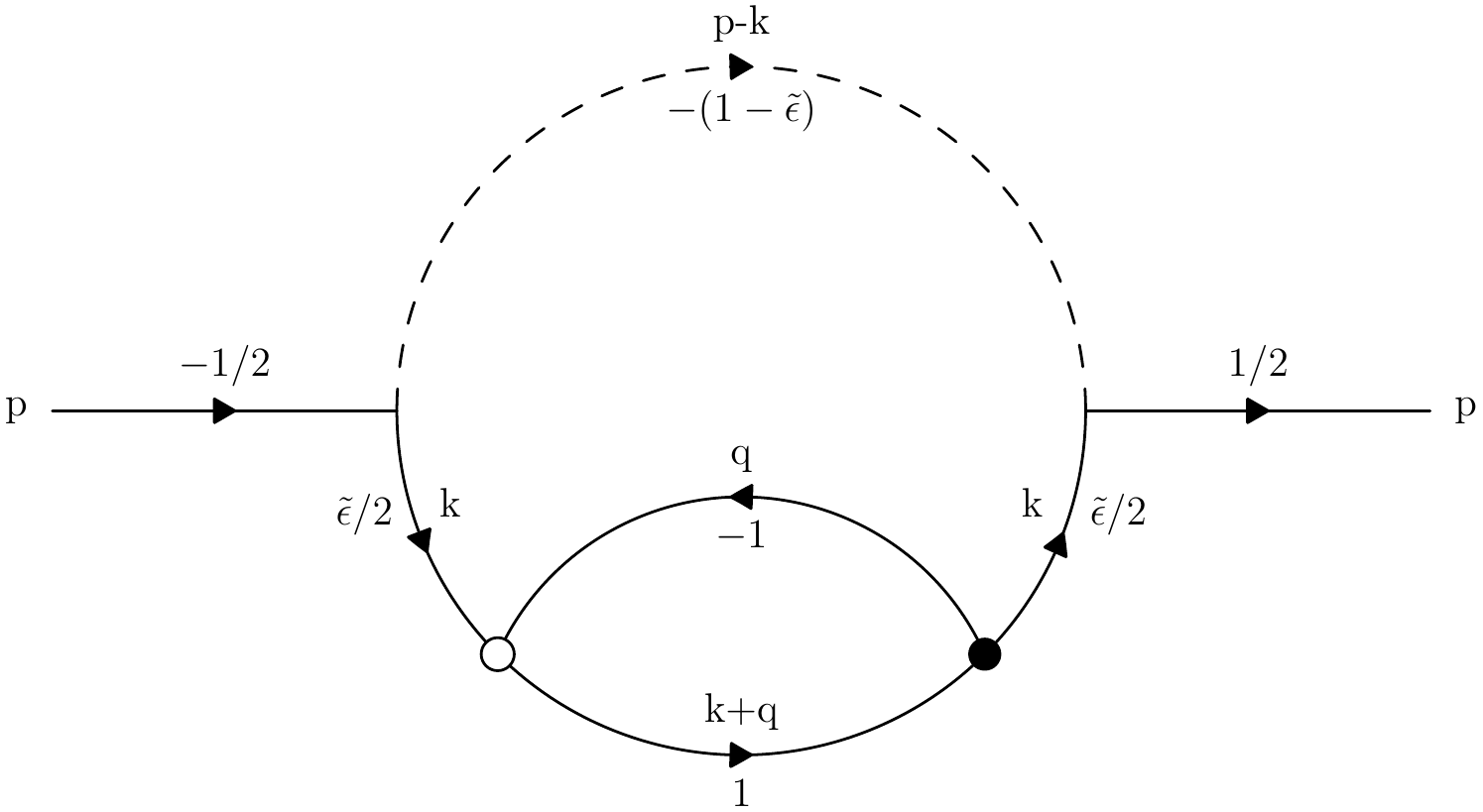}
  \caption{}
\end{subfigure}\\ 
\begin{subfigure}{.4\textwidth}
\centering
  \includegraphics[width=.9\textwidth, keepaspectratio]{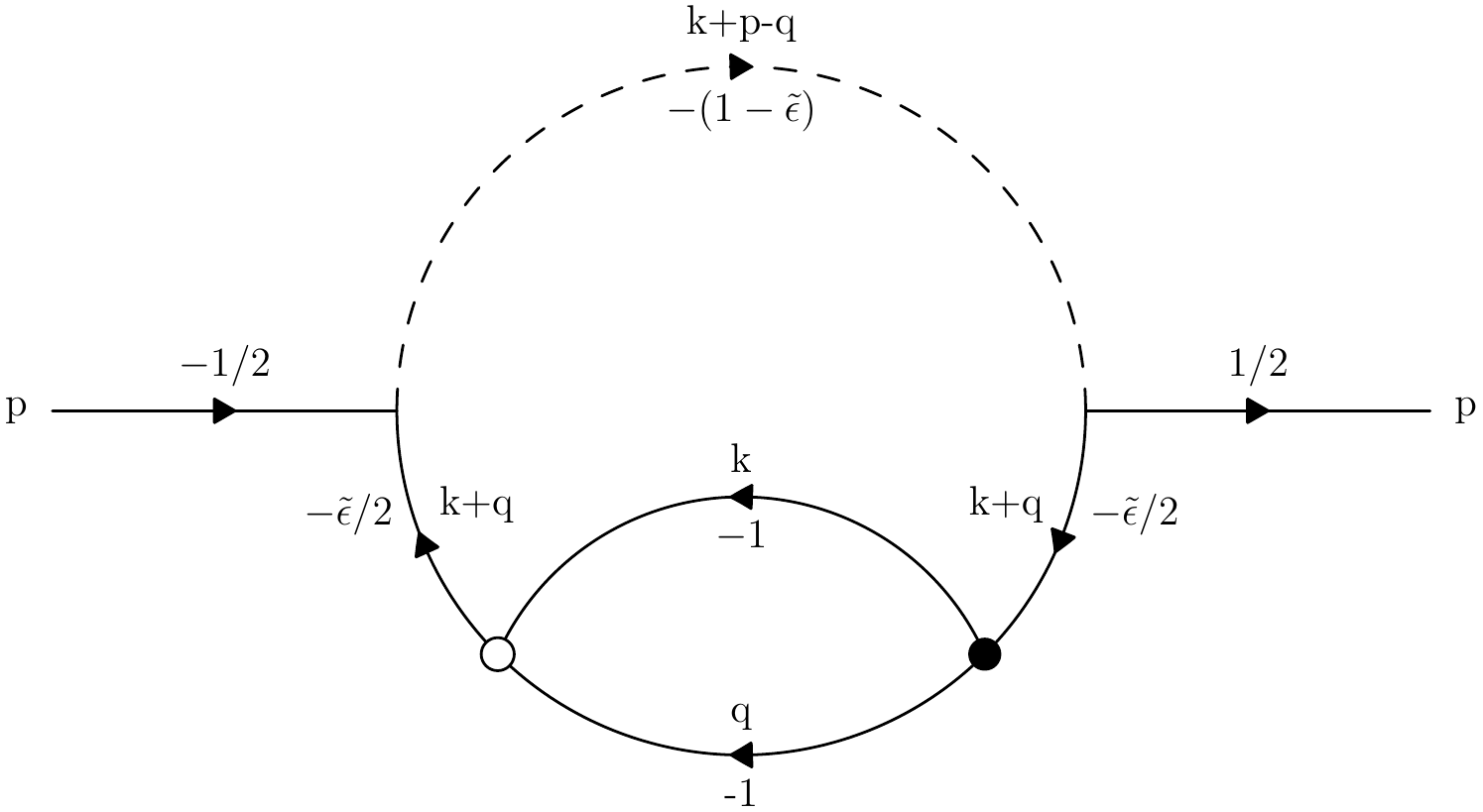}
  \caption{}
\end{subfigure}
\caption{Second type of 1PI corrections to the propagator from $\dD_4$.}
\label{D4F21PI2}
\end{figure}

There is also a second type of 1PI diagram, shown in figure \ref{D4F21PI2}.
The integrals again yield results of the form (\ref{logform}).

%%%%%%%%%%%%%%%%%%%%%%%%%%%%%%%%%%%%%%%%%%%%%%%%%%%%%%%%%%%%%%%

\subsection{$\bra {\mathbb L} \{ [ I^+ , U_{2f} ], \bar I^- \}   \ket {\mathbb L} $ terms}
Recall that 
\begin{equation}
U_{2f} = \int dz \left[ c_1 - c_2 \log(zK) \right] \rho^d(z) \check \rho^d(z)
\end{equation}
with $c_1=-2 (\gamma- \log 2), c_2 = -2$.  Let us now compute 
\begin{eqnarray} \nonumber
[I^+ , U_{2f} ] &=& {i \over 2 \sqrt 2} f^{abc} \int dx_1 dx_2 dz \sqrt{x_1 + x_2 \over x_1 x_2} \left[  c_1 - c_2 \log(zK)\right]   \\
\nonumber
&& \left\lbrace \delta(x_1 +x_2 -z) \rho^a(x_1) \rho^b(x_2) \check \rho^c(z) - 2 \delta(x-z) \rho^a(z) \rho^b(x_2) \check \rho^c(x_1 + x_2 )  \right\rbrace  \\
\nonumber
&=& {i c_2 f^{abc} \over 2 \sqrt 2}  \int dx_1 dx_2 \sqrt{x_1 + x_2 \over x_1 x_2 } \left[ - {c_1 \over c_2} + \log(K)- \log({x_1 + x_2 \over x_1 x_2})  \right] \rho^a(x_1) \rho^b(x_2) \check \rho^c(x_1 + x_2 ) \\
&=& {i c_2 f^{abc} \over 2 \sqrt 2 \bar \epsilon }  \int dx_1 dx_2 \left( {x_1 + x_2 \over x_1 x_2 } \right)^{{1 \over 2} - \bar \epsilon }  \rho^a(x_1) \rho^b(x_2) \check \rho^c(x_1 + x_2 )  + \Op(\bar \epsilon)
\end{eqnarray}
where $\bar \epsilon  = { 1 \over \log K - {c_1 \over c_2}}$.  Except for $\bar \epsilon$ corrections, this is almost same as $I^+$. From now on the computation is almost the same as that for $\dD_2 \sim \{ I^+ , \bar I^-\} $. In particular, we get the same diagrams as in section \ref{d2diagrams}, except that the momentum degree now get $\bar \epsilon$ corrections. In particular the diagram on left of Figure \ref{D2tree} still is regular. And the contribution for the diagram on the right is 
\begin{equation}
\sim \log(K) \int { du \over u^{1 - \tilde \epsilon } } \mathbb{J}^a(-u) \mathbb{J}^a(u)
\end{equation}
which obviously do not close the gap. 
The two fermion diagrams on the left of Figure \ref{D2loop} correspond to
\begin{equation}
N  \log(K) \int dp   \rho^a(p)  \check \rho^a(p) \ p^{1 - \bar \epsilon} \ \int_{\epsilon}^{p -\epsilon}  dk {1 \over k^{1 - \bar \epsilon} (p - k)^{1 - \bar \epsilon}}.
\end{equation} 
The diagram on the right gives  
\begin{equation} 
N \log(K) \int dp   \rho^a(p)  \check \rho^a(p) \ {1 \over p} \ \int_{\epsilon}^1 dk {(p+ k)^{1 - \bar \epsilon} \over k^{1 - \bar \epsilon} }. 
\end{equation}
Both contributions are again of the form (\ref{logform}).

%%%%%%%%%%%%%%%%%%%%%%%%%%%%%%%%%%%%%%%%%%%%%%%%%%
%%%%%%%%%%%%%%%%%%%%%%%%%%%%%%%%%%%%%%%%%%%%%%%%%%
%%%%%%%%%%%%%%%%%%%%%%%%%%%%%%%%%%%%%%%%%%%%%%%%%%

\end{document}